\documentclass[aps,twocolumn,preprintnumbers,amsmath,amssymb,superscriptaddress]{revtex4-2}
\UseRawInputEncoding
\usepackage[normalem]{ulem}

\usepackage{latexsym,amssymb,amsthm,amsmath,epsfig,braket} 
\usepackage[left=2.00cm, right=2.00cm, top=2.00cm, bottom=2.00cm]{geometry}
\usepackage{xcolor}
\usepackage{svg}
\usepackage[colorlinks, citecolor={blue!80!black}, urlcolor={blue!80!black}, linkcolor={red!50!black}]{hyperref}
\usepackage{booktabs}
\usepackage{subfigure}

\usepackage{hyperref}
\hypersetup{
    citecolor=red,
    colorlinks=true,
    linkcolor=blue,
    filecolor=blue,      
    urlcolor=blue,
}

\begin{document}
\title{Topological magnetotransport in modified-Haldane systems}
\author{A. Uzair}
\affiliation{National Centre for physics, Islamabad, 45320, Pakistan}
\author{Muzamil Shah}
\email{muzamil@qau.edu.pk}
\affiliation{Department of Physics, Quaid-I-Azam University Islamabad, 45320, Pakistan}
\author{Imtiaz Khan}
\affiliation{Department of Physics, Zhejiang Normal University, Jinhua, Zhejiang 321004, China}
\affiliation{Zhejiang Institute of Photoelectronics, Jinhua, Zhejiang 321004, China}
\author{Kashif Sabeeh}
\email{ksabeeh@qau.edu.pk}
\affiliation{Department of Physics, Quaid-I-Azam University Islamabad, 45320, Pakistan}
\date{\today}

\begin{abstract}
We present a theoretical study of quantum magneto-transport and magneto-optical (M-O) properties in modified-Haldane model; which is applicable to  diverse classes of two-dimensional (2D) quantum materials such as buckled Xene monolayers and transition metal dichalcogenide (TMDC) monolayers. By varying the staggered sublattice potential and intrinsic spin-orbit coupling, we identify distinct topological regimes and analyze their manifestations in the emergence of Landau levels, the evolution of the density of states, and the characteristics of M-O absorption spectra. Using the Kubo formalism, we compute the longitudinal and Hall M-O conductivities and show that inter-Landau-level (inter-LL) transitions produce characteristic resonance features that provide optical signatures of the underlying topological phases. Within this framework, we demonstrate electrically tunable topological phase transitions in buckled silicene. Extending our study to monolayer TMDCs, we show that inspite of large band gap, the spin-valley coupling provides a powerful tool for tailoring M-O absorption features across wide range of 2D materials. Collectively, these results underscore modified-Haldane-model materials as an ideal testbed for engineering quantum transport, with promising applications in topological photonics, valleytronic devices, and next-generation optoelectronics.
\end{abstract}

\maketitle

\section{Introduction}
The experimental realization of graphene in 2004 \cite{Novoselov2004}, followed by discovery of entire family of hexagonal 2D systems,  such as buckled 2D materials (silicene, germanene and stanene) \cite{ezawa2012topological,ezawa2013spin,Bampoulis202023quantum}, and various monolayer transition metal dichalcogenides (TMDCs) including $\mathrm{MoS_{2}}$ and $\mathrm{WSe_{2}}$ \cite{CHOI2017116} provides a wide tunable spectrum of electronic, optical, spintronic, optoelectronic and valleytronic properties, enabling a new generation of tailored quantum and nano-devices \cite{pesin2012spin,schaibley2016valleytronics}. The properties of these 2D quantum materials are fundamentally governed by the intrinsic symmetries of their shared hexagonal lattice structure, which is often depicted as an interconnected lattice of hexagons, exhibiting symmetry under time-reversal \(\mathcal{T}\) and inversion \(\mathcal{I}\) operations.

Breaking either $\mathcal{T}$ or $\mathcal{I}$ symmetry generates a band gap, transitioning from a semi-metal to a band insulator phase \cite{Haldane1988}. 
In graphene and buckled Xene monolayers, the preserved $\mathcal{T}$ and $\mathcal{I}$ symmetries give rise to pseudo-spin, while in layered TMDCs, and hexagonal boron nitride (hBN) nanosheets, preserved $\mathcal{T}$ and broken $\mathcal{I}$ symmetry, give rise to valley degree. Consequently, the interplay of spin and valley dynamics leads to various novel phenomena, characteristic of 2D hexagonal structures\cite{schaibley2016valleytronics,pesin2012spintronics,sasaki2008pseudospin}. 

The study of 2D systems under external magnetic field $B$ which explicitly breaks $\mathcal{T}$ and leads to Landau quantization serves as a premier testbed for exploring novel magneto-electronic and magneto-optoelectronic phenomena \cite{ando1982electronic,giuliani2005quantum,castro2009electronic,goerbig2011electronic}.

The absorption peaks in magneto-optical (M-O) conductivity arise from inter-LL transitions, which are optically permitted by the dipole selection rule \cite{novoselov2005two,zhang2005experimental}. 

These characteristic peaks in M-O spectra serve as sensitive spectroscopic probes for quantifying fundamental electronic properties like Fermi velocity, band gaps, and band structure features \cite{ando1982electronic,giuliani2005quantum,castro2009electronic,goerbig2011electronic,PhysRevLett.110.197402} and are studied using techniques such as magneto-transport, Hall measurements, magneto-infrared spectroscopy, and M-O methods \cite{PhysRevLett.125.046403,PhysRevB.102.235134,PhysRevB.105.195102}. Over the last decade, the M-O responses of 2D quantum materials including graphene, silicene, $\mathrm{MoS_{2}}$, and phosphorene have been extensively studied  \cite{castro2009electronic,tabert2013magneto,shah2019magneto,chu2014valley,tahir2015magneto, PhysRevB.107.235417,PhysRevB.109.165441,PhysRevB.101.045424}.  

F. D. M. Haldane in 1988 represented the first minimal lattice realization of a topological insulator (TI), demonstrating that a quantum Hall-like TI phase can arise without $B$ \cite{Haldane1988}. By introducing complex next-nearest-neighbor (NNN) hopping on a honeycomb lattice, he showed that $\mathcal{T}$ can be broken while preserving zero net magnetic flux through the unit cell, leading to chiral edge states. This seminal work provided the conceptual foundation for modern topological band theory and clarified that topology can emerge purely from band structure engineering rather than Landau quantization.
Building on this framework, Kane and Mele in 2005 generalized the Haldane model to include spin and intrinsic spin-orbit coupling (SOC), leading to the prediction of the quantum spin Hall effect \cite{KaneMele2005a,KaneMele2005b}. The Kane-Mele model consists of two time-reversed copies of the Haldane model for opposite spins, restoring $\mathcal{T}$ but yielding helical edge states protected against backscattering. Unlike the Haldane model, where topology is characterized by an integer Chern number, the Kane-Mele model introduced $Z_2$ topology and established the conceptual basis for practical realization of 2D TI in graphene-like systems.

More recently, modified-Haldane-type models have been proposed to capture additional symmetry-breaking mechanisms relevant to realistic 2D materials such as buckeld Xene \cite{PhysRevLett.110.197402,Bampoulis202023quantum,Khan:25,PhysRevB.101.205408,PhysRevLett.110.197402,tabert2013magneto,tahir2015magneto,Khan:25,ezawa2012spin,Ezawa_2012,Ezawa2012a,Ezawa2012,ezawa2012topological} and TMDCs \cite{PhysRevB.101.045424,Xiao2012,nature2014,PhysRevLett.113.077201,PhysRevB.109.165441}. These extensions incorporate staggered sublattice potentials, Rashba SOC, including an external magnetic field through both orbital and Zeeman coupling, exchange fields arising from magnetic substrates or proximity effects and application of strain. All the terms except the first two terms break $\mathcal{T}$ while the first two cause breaking of $\mathcal{I}$. Additionally, time-periodic perturbations such as off resonant circularly polarized light can also dynamically generate effective time-reversal-symmetry-breaking terms. All of these contributions lead to spin- and/or valley-dependent mass terms. In particular, the modified-Haldane model allows valley-contrasting Berry curvature and valley-dependent Chern numbers, enabling valley-polarized TI phases and tunable transitions between trivial and different topological regimes. Unlike the Kane-Mele model, where opposite spins carry opposite Chern numbers but total Chern number vanishes, the modified-Haldane framework permits simultaneous breaking of $\mathcal{T}$ and $\mathcal{I}$, leading to richer phase diagrams with spin- and valley-resolved topological responses. It is worthwhile to note that original Haldane model and the Kane-Mele model are special cases of modified-Haldane model, with inversion symmetry kept intact.

The primary motivation of this study is to use a single generic Hamiltonian as a unified, tuneable framework to describe M-O transport across diverse 2D quantum materials like Xenes and TMDCs. By adjusting material-specific parameters, this model captures different TI phases and transitions, allowing a direct comparison of how external perturbations influence systems ranging from trivial band insulators (BI)  to topological Dirac materials and valley-polarized semiconductors. Essentially, this generic framework has predictive power, enabling the identification of M-O signatures of both existing and hypothetical materials within this broad family, while clearly separating universal topological features from material-specific details. In this paper, we exploit this unified approach to systematically analyze the M-O response and highlight both the fundamental connections and distinctions across various 2D hexagonal modified-Haldane model materials. 

The modified-Haldane model, together with the original Haldane and Kane-Mele model, form the theoretical backbone for understanding topological magneto-transport phenomena in contemporary 2D quantum materials. From the M-O viewpoint, the modified-Haldane model provides a unified and highly tunable platform where inter-LL transitions, optical selection rules, and M-O Hall conductivities become valley- and spin-resolved. Unlike the Kane-Mele model, where opposite spin channels carry opposite Chern numbers and cancel in the total optical Hall response, the modified-Haldane framework permits a finite net optical Hall conductivity. In contrast to the original Haldane model, where $K$ and $K'$ valleys contribute symmetrically, the modified model predicts valley-asymmetric M-O spectra, enabling direct optical detection of topological phase transitions through the emergence or suppression of specific interband and inter-LL transitions.

The M-O response of the modified Haldane model carries direct and unambiguous signatures of the underlying TI phase. In the band-insulating (BI) regime, where the spin- and valley-dependent Dirac mass is positive, the zeroth Landau level is valley-band polarized: the valley $K(K')$ resides in the VB (CB). As a result, the inter-LL transition at the $K$ valley; $\Delta_{0\rightarrow1,K}$ is optically allowed. In contrast, in the TI regime, band inversion leads to spin-band polarization of the lowest LL (LLL), with $\downarrow(\uparrow)$ spins occupying the VB (CB), and the optically active transition shifts to $\Delta_{0\rightarrow1,\downarrow}$  in the $\downarrow$-spin sector.

Importantly, the Kramers pair $(K,\uparrow),(K',\downarrow)$ provides a robust optical fingerprint of the TI phase when monitoring the $\Delta_{0\rightarrow1}$ transition. The observation of the $\Delta_{0\rightarrow1,K,\uparrow}$ transition signifies a trivial band-insulating (BI) phase, whereas the activation of the $\Delta_{0\rightarrow1,K',\downarrow}$ transition indicates a TI phase. The complementary Kramer's pair exhibits identical behavior in both phases, therefore, does not serve as a phase discriminator. Transitions that are forbidden are Pauli blocked, resulting in vanishing optical conductivity due to the occupation factors in the Kubo formalism.

We employ a generic modified-Haldane model Hamiltonian and calculate the LL dispersion relation.   Initially, we scrutinize these LLs within both topological and trivial systems, providing insights into the distinct spectral signatures characterizing each phase. We thoroughly analyze LL formation and the subsequent evolution of the density of states (DOS). The DOS are computed for the valleys in TI and BI regimes, highlighting the influence of topological phase transitions. In the TI phase, the M-O response is inherently valley-selective, a sharp contrast to the valley-independent behavior observed in the BI phase. The field-induced shift of the lowest LL (LLL) from the valence band (VB) to the conduction band (CB) results in Pauli blocking of a certain transition, leaving a distinct imprint on the optical conductivity. We observed that upward optical transitions originating from the LLL are suppressed by Pauli blocking, whereas downward transitions from higher LLs into the LLL remain allowed. 

Subsequently, we broaden our investigation to encompass the realm of buckled 2D monolayer material. We explore potential M-O transitions between quantized LLs, governed by electric-dipole selection rules. With the Fermi level set at zero energy, only inter-band transitions are discussed. Our analysis reveals how tuning parameters defining SOC and staggered sublattice potential shifts the valley and spin polarized LLLs across the Fermi level, thereby changing the system's topology and directly affecting the M-O response. We demonstrate that the M-O response of buckled Xene monolayers offers a clear and robust probe of their underlying topological character.  Consequently, different TI phases of silicene are characterized by excitation peaks at well-separated energies, providing unambiguous optical signatures of phase transitions. The accompanying valley-dependent Berry curvature further emphasizes the strong  spin and valley coupled topology. Lastly, our exploration delves into the spin and valley-polarized M-O response within monolayer TMDCs. Our findings highlight the remarkable sensitivity of these M-O effects to spin and valley indices but ignoring the band dependence of effective masses inherent in these materials.

This paper is organized as follows. In Sec.\ref{energystate}, we first outline the essential concepts of the modified Haldane Hamiltonian, and then derive the eigen-energies and eigen-states in presence of magnetic field $B$. We present full theoretical framework of the formation of Landau levels and evolution of the DOS. Explicit expressions of Berry curvature and M-O conductivity are also presented using Kubo formalism. In Sec.\ref{moresponse}, quantum phase transitions driven by the competition between staggered sublattice potential and SOC are identified through M-O conductivity and the associated allowed transitions. The LL spectrum, DOS and M-O responses of buckled Xene monolayers and TMDCs are discussed in separate subsections. Finally, Sec. \ref{con} concludes the key findings of this work

\section{Theoretical Framework for Magneto-Optical Response in Modified-Haldane MODEL}\label{energystate}
\subsection{Modified-Haldane Model Hamiltonian}
 While the original Haldane model is governed exclusively by complex and chiral next-nearest-neighbor (NNN) hopping, the modified Haldane model incorporates spin-dependence of NNN hopping and an additional inversion symmetry breaking term. Including both spins makes the model invariant under $\mathcal{T}$; just like Kane-Mele and can be applied to real 2D hexagons like silicene which have significant SOC. However, the application of perpendicular electric field can be used to break $\mathcal{I}$ in silicene which is intrinsically broken in TMDCs. It is the relative strength and competition between these two mechanisms that dictates the realization of distinct topological phases. Specifically, we consider a generic tight-binding Hamiltonian, featuring simultaneous breaking of time-reversal and inversion symmetries \cite{Haldane1988,vanderbilt2018berry,pratama2020circular}. The modified Haldane model Hamiltonian is given as:
\begin{equation}\label{a1}
    \hat{H}=-t_{1} \sum_{(i, j)} a_{i}^{\dagger} a_{j}+t_{2} \sum_{\langle(i, j)\rangle} e^{i v_{ij}\phi} a_{i}^{\dagger} a_{j}+\mathcal{M} \sum_{i} \chi_{i} a_{i}^{\dagger} a_{i},
\end{equation}
 where, $a_{i}^{\dagger}(a_{i})$ denotes the fermionic creation(annihilation) operator at the lattice site $i$. The hopping amplitudes $t_{1}$ and $t_{2}$ correspond to nearest-neighbor (NN) and  next-nearest-neighbor (NNN) interactions respectively.

The complex part of the NNN hopping term in Eq.~\eqref{a1} explicitly breaks time-reversal symmetry $\mathcal{T}$ for any nonzero phase angle $\phi$ appearing in the exponential factor $e^{i\nu_{ij}\phi}$. This phase is acquired by an electron when it moves in the presence of a vector potential $\vec{A}$. This phase, called Peierls phase is given by $\phi_{ij}=\frac{e}{\hbar}\int_i^j \vec{A}\cdot d\vec{l},$ where $d\vec{l}$ denotes the bond element along the hopping path. In hexagonal lattice systems, complex NNN hopping effectively mimics such a phase, and thereby generates an effective vector potential. The real part of the NNN hopping merely shifts the energy spectrum about $\mu=0$ and thereby breaks particle-hole symmetry.

For a closed hopping path, the accumulated phase $\phi$ is related to the magnetic flux enclosed by the loop $\Phi$ via $\phi=\frac{e}{\hbar}\oint \vec{A}\cdot d\vec{l}=\frac{e}{\hbar}\Phi$. Thus, whenever a finite flux $\Phi$ is enclosed, the electron acquires a phase $\phi$. Haldane defined hexagonal plaquette such that the net flux through hexagonal unit cell vanishes but staggered flux in triangular sublattices survives. In natural units $(e=\hbar=1)$, the phase $\phi$ and the enclosed flux $\Phi$ become numerically equivalent and are therefore often used interchangeably in the literature.

The factor $\nu_{ij}$ encodes the chirality of the NNN hopping. In the honeycomb lattice, the triangular plaquettes formed by the $A$ and $B$ sublattices have opposite orientations. Consequently, electrons circulating around these triangular loops experience equal and opposite effective fluxes $\pm\Phi$. Specifically, $\nu_{ij}=+1~(-1)$ corresponds to clockwise (anticlockwise) NNN hopping paths, as illustrated in Fig.~\ref{phase}(a).

The last term represents the staggered on-site potential between the $A$ and $B$ sublattices, with $\chi_{i}=+1(-1)$ for the $A(B)$ sublattice. This term breaks the $A$ and $B$ sublattice inversion symmetry $\mathcal{I}$, and opens a band gap of magnitude $2M$ at the $K$ and $K^{\prime}$ valleys of the Brillouin zone as presented in Fig. \ref{phase}(b). Throughout this work, we consider sufficiently large systems such that edge effects can be neglected. Upon Fourier transformation, the Hamiltonian can be expressed in momentum space, yielding the corresponding low-energy effective Hamiltonian near the $K$ and $K^{\prime}$ points, which is block diagonal in $2*2$ matrices labeled by valley and spin  \cite{cooper2019topological,ren2016topological}:
\begin{eqnarray} \label{a2}
\hat{\mathcal{H}}(k)=-\lambda_s\sigma_{0}+\hbar v_{F}\left(\tau k_{x} \hat{\sigma}_{x}+k_{y} \hat{\sigma}_{y}\right)+\Delta_{\tau,s} \hat{\sigma}_{z},
\end{eqnarray}
where, $\lambda_s=3 t_{2} \cos \phi$ is the band gap and depends on the strength of NNN hopping $t_2$ for 2D hexagonal materials. 
The second term corresponds to the familiar low-energy graphene-like Dirac Hamiltonian that captures the physics of massless Dirac fermions near the Dirac points. Here $k_{x,y}$ denote the crystal momentum components, and the Fermi velocity is given by $v_F=\frac{\sqrt{3}\,a\,t_1}{2\hbar},$where $a$ is the lattice constant and $t_1$  is the nearest-neighbour hopping amplitude. The index $\tau=+1(-1)$ labels the $K\,(K')$ valleys in momentum space. The operators $\sigma_{x,y,z}$ are the Pauli matrices acting in the sublattice (pseudospin) space.

The last term in the Hamiltonian corresponds to the effective Dirac mass, $\Delta_{\tau,s}=M-\tau3\sqrt{3}\,t_2\sin\phi$. To analyze the TI (BI) regimes, one typically considers the imaginary part of the hopping, characterized by $\sin\phi$, to be maximal (minimal). To study the BI regime, the Haldane mass term proportional to $\sin\phi$ is taken to vanish ($\phi=0$). In contrast, the TI regime is obtained when $\sin\phi\neq0$ i.e. complex term survives and TRS breaking NNN hopping dominates inversion symmetry breaking staggered sublattice potential $M$.

In the original Haldane model, the NNN hopping is complex and chiral but spin-independent, leading to breaking of $\mathcal{T}$. In contrast, in materials with significant SOC, effective complex NNN hopping arises from intrinsic SOC and is spin dependent while preserving $\mathcal{T}$. Importantly, NNN hopping connects sites within the same sublattice and can therefore generate a sublattice-preserving mass term, whereas NN hopping connects different sublattices and cannot produce such a term. Consequently, SOC naturally emerges through same-sublattice virtual hopping processes, making NNN hopping the lowest-order symmetry-allowed contribution.

In this picture, electrons with opposite spins acquire equal and opposite effective phases ($\pm \phi$) when traversing triangular loops, corresponding to opposite effective fluxes ($\pm \Phi$). This spin-dependent NNN hopping is formally analogous to the intrinsic SOC term in the Kane-Mele model, where the spin degree of freedom is encoded through $\sin\phi$, the orbital chirality through $\nu_{ij}$, and the resulting SOC-induced gap is given by $\Delta_{so} = 3\sqrt{3}t_2$. The real spin degree of freedom is not explicitly written in the Dirac term but enters through the complex NNN hopping phase $\phi$ as explained above.

Note that the $\cos\phi$ term breaks particle-hole symmetry and produces an equal energy shift at both valleys and on both sublattices. For systems  with approximate particle-hole symmetry such as silicene, the Haldane mass is often maximized by choosing $\phi=\pm\pi/2$, which suppresses the contribution of $\cos\phi$ and minimizes the energy shift. For particle-hole asymmetric systems, such as TMDCs,, a general value of $\phi$ is considered as this symmetry is intrinsically broken in TMDCs as a result of the distinct orbital character of the CB and VB, unequal effective masses, and the asymmetric action of SOC, which predominantly affects the VB. As a result, the electronic spectrum lacks symmetry between positive and negative energies.

In the presence of a static and uniform magnetic field perpendicular to the plane of 2D hexagonal materials, Hamiltonian modifies as 
\begin{eqnarray}\label{a3}
    H=\begin{pmatrix}
-\lambda_s+\Delta_{\tau,s} &\hbar v_{F}\left(\tau \pi_{x} -i \pi_{y}\right) \\
\hbar v_{F}\left(\tau \pi_{x} +i \pi_{y}\right)  & -\lambda_s-\Delta_{\tau,s}
\end{pmatrix}.
\end{eqnarray}
The magnetic vector potential for Landau guage is $A=(-yB,0,0)$, where $\hbar\pi=\hbar k-e A$. The Hamiltonian changes to
\begin{widetext}
    \begin{eqnarray}\label{a4}
    H=\begin{pmatrix}
-\lambda_s+\Delta_{\tau,s} &\hbar v_{F}\left(\tau (k_{x}-eBy/\hbar) -i k_{y}\right)\\
\hbar v_{F}\left(\tau (k_{x}-eBy/\hbar) +i k_{y}\right)  & -\lambda_s-\Delta_{\tau,s}
\end{pmatrix}.
\end{eqnarray}
\end{widetext}

\subsection{Landau Levels}
 A perpendicular magnetic field quantizes the energy spectrum into discrete Landau levels (LLs) which  form the fundamental basis for understanding a range of magneto-transport and M-O phenomena \cite{klitzing1986quantized,stern1967properties,chu2014valley,yuan2018chiral,li2013magneto}. Diagonalizing the Hamiltonian in Eq. \eqref{a4} gives the eigenvalues
\begin{align}
    E_{n,\eta}^{\tau,s} &= 
\begin{cases}
-\lambda_s+\eta \sqrt{ 2v_F^2\hbar eB|n| + \Delta_{\tau,s}^2} & n \neq 0,\\
-\lambda_s-\tau\Delta_{\tau,s} & n=0.\label{energy}
\end{cases} 
\end{align}
This depicts the formation of unequally spaced LLs that scale as $\sqrt{nB}$. $E\propto \sqrt{n} $  implying that the spacing decreases as $n$ increases.  This contrasts starkly with conventional 2D systems where the LL energy scales linearly ($E_n \propto nB$) . The lowest Landau level (LLL), i.e., level with the lowest $n,(n=0)$ is not always pinned at zero energy as in graphene but depends on the valley and spin of the electron. For different values of $t_2, M \text{ and } \phi$ hidden in $\lambda_s$ and $\Delta_{\tau,s}$, the position of the LLL changes and new TI phases emerge as explained later. It is worthy to note that position of LLL determines the topology of the phase. The eigen functions in terms of spinors for $K$ and $K'$ valley are given as:
\begin{align}
| \psi_{n, \eta}^{\tau, s}\rangle _{\tau=1}&=
\begin{pmatrix}
{-i \mathcal{A}_{n, \eta}^{\tau, s}|\phi_{n-1}\rangle}\\{\mathcal{B}_{n, \eta}^{\tau, s}|\phi_{n}\rangle}
\end{pmatrix},
\end{align}
\begin{align}
| \psi_{n, \eta}^{\tau, s}\rangle_{\tau=-1} &=
\begin{pmatrix}{-i \mathcal{A}_{n, \eta}^{\tau, s}|\phi_{n}\rangle}\\
{\mathcal{B}_{n, \eta}^{\tau, s}|\phi_{n-1}\rangle}
\end{pmatrix},
\end{align}
where $|\phi_n\rangle$  is the orthonormal Fock state of harmonic oscillator. The spinor has its origin in $A$ and $B$ sublattice basis. The coefficients $\mathcal{A}_{n, \eta}^{\tau, s}$ and $\mathcal{B}_{n, \eta}^{\tau, s}$ describe the probability amplitude of finding electrons at lattice sites $A$ and $B$ respectively and are given as: 
\begin{align} 
\mathcal{A}_{n, \eta}^{\tau, s} &= 
\begin{cases}
\sqrt{\dfrac{|E_{n,\eta}^{\tau,s}|+\eta\Delta_{\tau,s}}{2|E_{n,\eta}^{\tau,s}|}} & n \neq 0,\\
\dfrac{1-\tau}{2} & n=0,
\end{cases}  \\
\mathcal{B}_{n, \eta}^{\tau, s} &= 
\begin{cases}
\sqrt{\dfrac{|E_{n,\eta}^{\tau,s}|-\eta\Delta_{\tau,s}}{2|E_{n,\eta}^{\tau,s}|}} & n \neq 0,\\
\dfrac{1+\tau}{2} & n=0.
\end{cases}
\end{align}
Its worth-mentioning here that LLLs at $K$ valley are restricted to sublattice $B$, whereas those at $K'$ valley have a non-vanishing value only on the sublattice $A$.

\subsection{Density of States}\label{dosapp}
The electronic density of states (DOS) of modified-Haldane materials in $K$ and $K'$ valleys can be calculated from Eq. \eqref{energy}. The DOS describes the number of available electronic states per unit energy that can be occupied by an electron and is given by \cite{tabert2013magneto}
\begin{equation}\label{a8}
D(E)=D_0 \sum_{ \tau,\eta=\pm} \sum_{n=0}^{\infty} \delta[E-E^{\tau,s}_{n,\eta}].
\end{equation}
In the presence of magnetic field $B$, time-reversal symmetry is broken and the electronic spectrum is quantized into Landau levels, rendering the wave vector
$k$ no longer a good quantum number. Consequently, the DOS becomes independent of $k$. The prefactor $D_0$ accounts for the spin and valley degeneracy of each LL and is given by $D_0=e B /(2 \pi \hbar c)$. For simplicity we set $c=1$. In numerical calculations, Dirac-$\delta$-function is replaced by Lorentzian function to incorporate Landau-level broadening in terms of decay rate $\Gamma$.

\subsection{Berry Curvature}
Berry curvature plays a role analogous to that of a gauge-field in electrodynamics: it is a gauge-invariant quantity and therefore directly observable. Beyond governing semiclassical electron dynamics, Berry curvature influences transport phenomena, modifies the electronic DOS in phase space, and affects the thermodynamic properties of crystalline solids \cite{xiao2010berry}.
\begin{widetext}
\begin{equation}\label{a9}
\Omega_{\mu \nu}^n=\frac{i \sum_{n \neq m}\langle \psi_{n, \eta}^{\tau, s}|\frac{\partial H}{\partial R^\mu}|\psi_{m,\eta'}^{\tau, s}\rangle\langle \psi_{m, \eta'}^{\tau, s}\frac{\partial H}{\partial R^\nu}|\psi_{n, \eta}^{\tau, s}\rangle-(\nu \leftrightarrow \mu)}{\left(E_n-E_{m}\right)^2} .
\end{equation}
\end{widetext}
 Physically, Berry curvature acts as an effective magnetic field in momentum space, deflecting charge carriers transversely even in the absence of an external magnetic field. As a result, each electronic state acquires an intrinsic Hall response.
 
 In Landau-quantized systems with $n\neq0$, the structure of the Berry curvature is governed by the mathematical properties of Hermite polynomials. In particular, their orthogonality relations impose conditions on the allowed matrix elements, leading to characteristic features in the Berry curvature.
 \begin{widetext}
\begin{equation}\label{a72}
 \Omega_{n}^{\tau,s}=\frac{\tau \hbar^2 v_{F}^2 \sqrt{2v_F^2\hbar eB|n|}}{2 \sqrt{2v_F^2\hbar eB|n|+\Delta_{\tau,s}^2}} 
\left(\frac{1}{\sqrt{2v_F^2\hbar eB|n|+\Delta_{\tau,s}^2}+\sqrt{2v_F^2\hbar eB|n+1|+\Delta_{\tau,s}^2}}\right)^2.
\end{equation}
\subsection{Magneto-optical conductivity}\label{mocon}
The M-O conductivity of modified-Haldane-type materials is calculated analytically within linear response theory using the Kubo formalism, expressed in terms of the system’s eigenvalues and eigenfunctions. 
The conductivity tensor can be written as
\begin{equation}\label{a10}
	\sigma_{\mu\nu}(\omega)
	=\frac{i\hbar}{2\pi l_{B}^{2}}\sum_{s,\tau, \eta,\eta'}\sum_{m,n}\frac{f_{n, \eta}^{\tau, s}-f_{m, \eta'}^{\tau, s}}{E_{n, \eta}^{\tau, s}-E_{m, \eta'}^{\tau, s}}\thickspace
	\frac{\langle \psi_{n, \eta}^{\tau, s}|\hat{j}_{\mu}|\psi_{m,\eta'}^{\tau, s}\rangle\langle \psi_{m, \eta'}^{\tau, s}|\hat{j}_{\nu}|\psi_{n, \eta}^{\tau, s}\rangle}{\hbar \omega-(E_{n, \eta}^{\tau, s}-E_{m, \eta'}^{\tau, s})+i\Gamma},
	\end{equation}
where the Fermi--Dirac distribution function is given by
\begin{equation}
f_{n, \eta}^{\tau, s} = \frac{1}{1 + \exp\left(\frac{E_{n, \eta}^{\tau, s} - \mu_F}{k_B T}\right)},
\end{equation}
which gives the probability that a state with energy $E_{n, \eta}^{\tau, s}$ is occupied at temperature $T$ and chemical potential $\mu_F$. \( E_{n,\eta}^{\tau,s} \) represents the energy of the \( nth \) Landau Level, \( \Gamma \) characterizes the scattering rate related to transport phenomenon and accounts for the LL-broadening.  The current operator is given as \( \hat{j}_\mu = e v_F \hat{\sigma}_\mu \), where \( \hat{\sigma}_\mu \) are the Pauli matrices, and the index \( \mu \in \{x, y, z\} \) represents the spatial direction.  Here \( l_B = \sqrt{\hbar / eB} \) quantifies the magnetic length. The longitudinal M-O conductivity can be obtained as \cite{tabert2013magneto,shah2019magneto}
\begin{equation}\label{a15}
\begin{aligned}
\sigma_{x x}(\omega)=\frac{i \hbar e^2 v_F^2}{2 \pi l_B^2} \sum_{\tau, s= \pm} \sum_{n, m=0}^{\infty} \sum_{\eta, \eta^{\prime}= \pm} & \frac{\Theta\left(\mu_{F}-E_{m, \eta'}^{\tau, s}\right)-\Theta\left(\mu_{F}-E_{n, \eta}^{\tau, s}\right)}{E_{n,\eta}^{\tau, s}-E_{m, \eta'}^{\tau, s}} \\
& \times \frac{\left(\mathcal{A}_{m, \eta'}^{\tau, s} \mathcal{B}_{n, \eta}^{\tau, s}\right)^2 \delta_{m-\tau, n}+\left(\mathcal{B}_{m, \eta'}^{\tau, s} \mathcal{A}_{n, \eta}^{\tau, s}\right)^2 \delta_{m+\tau, n}}{\hbar \omega+E_{m, \eta'}^{\tau, s}-E_{n,\eta}^{\tau, s}+i \Gamma} .
\end{aligned}
\end{equation}
and transverse M-O conductivity can be written as
\begin{equation}\label{a15a}
\begin{aligned}
\sigma_{x y}(\omega)=\frac{i \hbar e^2 v_F^2}{2 \pi l_B^2} \sum_{\tau, s= \pm} \sum_{n, m=0}^{\infty} \sum_{\eta, \eta^{\prime}= \pm} & \frac{\Theta\left(\mu_{F}-E_{m, \eta'}^{\tau, s}\right)-\Theta\left(\mu_{F}-E_{n, \eta}^{\tau, s}\right)}{E_{n,\eta}^{\tau, s}-E_{m, \eta'}^{\tau, s}} \\
& \times i\tau\frac{\left(\mathcal{A}_{m, \eta'}^{\tau, s} \mathcal{B}_{n, \eta}^{\tau, s}\right)^2 \delta_{m-\tau, n}-\left(\mathcal{B}_{m, \eta'}^{\tau, s} \mathcal{A}_{n, \eta}^{\tau, s}\right)^2 \delta_{m+\tau, n}}{\hbar \omega+E_{m, \eta'}^{\tau, s}-E_{n,\eta}^{\tau, s}+i \Gamma} .
\end{aligned}
\end{equation}
These analytical expressions characterize the longitudinal and transverse(Hall) M-O conductivities of the graphene-family under applied electric and magnetic fields. The conductivity scale $\sigma_0=e^2/(4\hbar)$ corresponds to the universal M-O conductivity of graphene. The Kronecker $\delta$-functions enforce electric-dipole selection rules, allowing optical transitions only between LLs with indices $m$ and $n=m\pm1$, consistent with harmonic-oscillator ladder operators. The Heaviside step functions $\Theta(E_n-\mu_F$), originating from the Fermi-Dirac distribution, restrict allowed optical transitions to those that cross the Fermi level, thereby incorporating Pauli blocking effects 
In addition, real-spin conservation imposes a strict selection rule, rendering transitions between opposite-spin states $(s=\pm1)$ spin forbidden. The real and imaginary parts of longitudinal conductivity can be obtained as 
\begin{equation}\label{a16}
\begin{aligned}
\operatorname{Re} & \bigg(\frac{\sigma_{x x}(\omega)}{\sigma_0}\bigg)=\frac{2 v_F^2 \hbar e B}{\pi} \sum_{\tau, s= \pm} \sum_{n, m=0}^{\infty} \sum_{\eta, \eta^{\prime}= \pm} \frac{\Theta\left(\mu_{F}-E_{m, \eta'}^{\tau, s}\right)-\Theta\left(\mu_{F}-E_{n, \eta}^{\tau, s}\right)}{E_{n, \eta}^{\tau, s}-E_{m, \eta'}^{\tau, s}} \\
& \times\left[\left(\mathcal{A}_{m, \eta'}^{\tau, s} \mathcal{B}_{n,\eta}^{\tau, s}\right)^2 \delta_{n, m-\tau}+\left(\mathcal{B}_{m, \eta'}^{\tau, s} \mathcal{A}_{n,\eta}^{\tau, s}\right)^2 \delta_{n, m+\tau}\right] \frac{\Gamma}{\Gamma^2+\left(\hbar \omega+E_{m, \eta'}^{\tau, s}-E_{n,\eta}^{\tau, s}\right)^2},
\end{aligned}
\end{equation}
and 
\begin{equation}\label{a17}
\begin{aligned}
\operatorname{Im} & \bigg(\frac{\sigma_{x x}(\omega)}{\sigma_0}\bigg)=\frac{2 v_F^2 \hbar e B}{\pi} \sum_{\tau, s= \pm} \sum_{n, m=0}^{\infty} \sum_{\eta, \eta^{\prime}= \pm} \frac{\Theta\left(\mu_{F}-E_{m, \eta'}^{\tau, s}\right)-\Theta\left(\mu_{F}-E_{n, \eta}^{\tau, s}\right)}{E_{n, \eta}^{\tau, s}-E_{m, \eta'}^{\tau, s}} \\
& \times\left[\left(\mathcal{A}_{m, \eta'}^{\tau, s} \mathcal{B}_{n,\eta}^{\tau, s}\right)^2 \delta_{n, m-\tau}+\left(\mathcal{B}_{m, \eta'}^{\tau, s} \mathcal{A}_{n,\eta}^{\tau, s}\right)^2 \delta_{n, m+\tau}\right] \frac{\hbar \omega+E_{m, \eta'}^{\tau, s}-E_{n,\eta}^{\tau, s}}{\Gamma^2+\left(\hbar \omega+E_{m, \eta'}^{\tau, s}-E_{n,\eta}^{\tau, s}\right)^2},
\end{aligned}
\end{equation}
 
Analogously, one can derive explicit expressions for the real and imaginary components of the transverse (Hall) conductivity, \cite{shah2019magneto}

\begin{equation}\label{a18}
\begin{aligned}
\operatorname{Re} & \bigg(\frac{\sigma_{x y}(\omega)}{\sigma_0}\bigg)=\frac{2 v_F^2 \hbar e B}{\pi} \sum_{\tau, s= \pm} \sum_{n, m=0}^{\infty} \sum_{\eta, \eta^{\prime}= \pm} \frac{\Theta\left(\mu_{F}-E_{m, \eta'}^{\tau, s}\right)-\Theta\left(\mu_{F}-E_{n, \eta}^{\tau, s}\right)}{E_{n, \eta}^{\tau, s}-E_{m, \eta'}^{\tau, s}} \\
& \times\left[\left(\mathcal{A}_{m, \eta'}^{\tau, s} \mathcal{B}_{n,\eta}^{\tau, s}\right)^2 \delta_{n, m-\tau}-\left(\mathcal{B}_{m, \eta'}^{\tau, s} \mathcal{A}_{n,\eta}^{\tau, s}\right)^2 \delta_{n, m+\tau}\right] \frac{\hbar \omega+E_{m, \eta'}^{\tau, s}-E_{n,\eta}^{\tau, s}}{\Gamma^2+\left(\hbar \omega+E_{m, \eta'}^{\tau, s}-E_{n,\eta}^{\tau, s}\right)^2},
\end{aligned}
\end{equation}

and
\begin{equation}\label{a19}
\begin{aligned}
\operatorname{Im} & \bigg(\frac{\sigma_{x y}(\omega)}{\sigma_0}\bigg)=-\frac{2 v_F^2 \hbar e B}{\pi} \sum_{\tau, s= \pm} \sum_{n, m=0}^{\infty} \sum_{\eta, \eta^{\prime}= \pm} \frac{\Theta\left(\mu_{F}-E_{m, \eta'}^{\tau, s}\right)-\Theta\left(\mu_{F}-E_{n, \eta}^{\tau, s}\right)}{E_{n, \eta}^{\tau, s}-E_{m, \eta'}^{\tau, s}} \\
& \times\left[\left(\mathcal{A}_{m, \eta'}^{\tau, s} \mathcal{B}_{n,\eta}^{\tau, s}\right)^2 \delta_{n, m-\tau}-\left(\mathcal{B}_{m, \eta'}^{\tau, s} \mathcal{A}_{n,\eta}^{\tau, s}\right)^2 \delta_{n, m+\tau}\right] \frac{\Gamma}{\Gamma^2+\left(\hbar \omega+E_{m, \eta'}^{\tau, s}-E_{n,\eta}^{\tau, s}\right)^2}.
\end{aligned}
\end{equation}
In the limiting case $\Delta_{so}=M=\lambda=0$, our formalism correctly recovers the unconventional Hall response of graphene \cite{PhysRevLett.98.157402,gusynin2005unconventional}.
\end{widetext}

\section{Landau Quantization and Magneto-Optical Response of Buckled Xenes and TMDCs}\label{moresponse}

\subsection{Trivial and Topological}

\begin{figure}[ht!]
	\centering		
\includegraphics[width=0.850\linewidth]{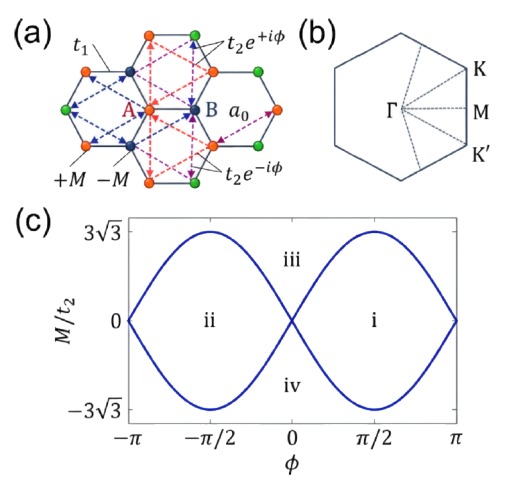}
\caption{(a) Schematic of the modified-Haldane model on the 2D hexagonal lattice consisting of $A$ and $B$ sublattices with lattice constant $a_0$. The nearest neighbor(NN) hopping amplitude is denoted by $t_1$ while the complex next-nearest neighbor (NNN) hopping is given by $t_2 e^{+i \phi}\left(t_2 e^{-i \phi}\right)$ for the clockwise (anticlockwise) hopping paths. The on-site sublattice potential is $+M(-M)$ for $A(B)$ sublattice. (b) The first Brillouin zone of the 2D hexagonal lattice indicating high-symmetry points. (c) Phase diagram of the modified-Haldane model, where regions (i) and (ii) correspond to the TI phases, while regions (iii) and (iv) represent trivial BI phase.}
\label{phase}
\end{figure}

The modified-Haldane model hosts distinct TI phases  governed by onsite potential difference $M$ between the $A$ and $B$ sublattices, the strength of NNN interactions parametrized by hopping amplitude $t_2$, and time reversal symmetry breaking determined by phase $\phi$.  An additional gap contribution arising from the term $\lambda$ i.e. real part of NNN hopping breaks particle-hole symmetry and modifies LL spectra in terms of energy shift only. The trivial case is defined by vanishing Haldane mass i.e. $\sin\phi=0$, so only the real NNN hopping survives. For non-zero $\phi$, the phase will be determined by relatively larger value among $M$ and $t_2$. It is worth mentioning that the topological phase transition occurs when the relative amplitude of the essential parameters $M$ and $t_2$ varies. When $ |M/t_2|$ is smaller than $3\sqrt 3|\sin\phi|$, system is in TI phase. On the other hand, if $ |M/t_2|$ is greater than $3\sqrt 3|\sin\phi|$, then it becomes a trivial BI as shown in Fig. \ref{phase} (c). In TI phase, the complex contribution in NNN hopping is maximized by setting $\phi=\pi/2$ . The band gap in this phase will be determined solely by the inversion asymmetry $M$ or the effective SOC $3 \sqrt 3 t_2$ which implies this gap will depend on spin and valley degrees of freedom. 

\begin{figure*}[ht!]
	\centering\scalebox{1}{
    \begin{tabular}{cc}
	\includegraphics[width=0.450\linewidth]{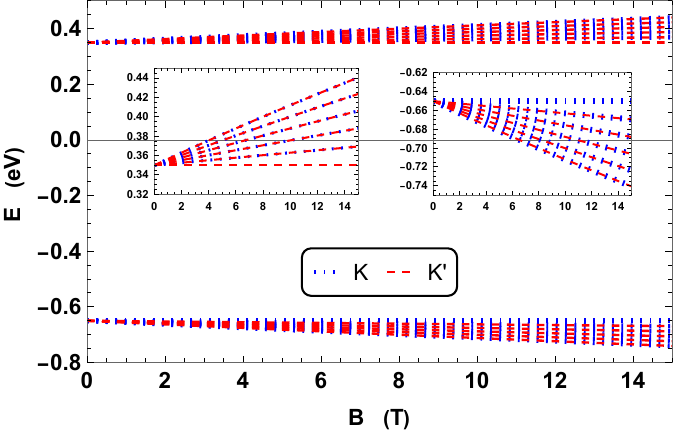}&
    \includegraphics[width=0.450\linewidth]{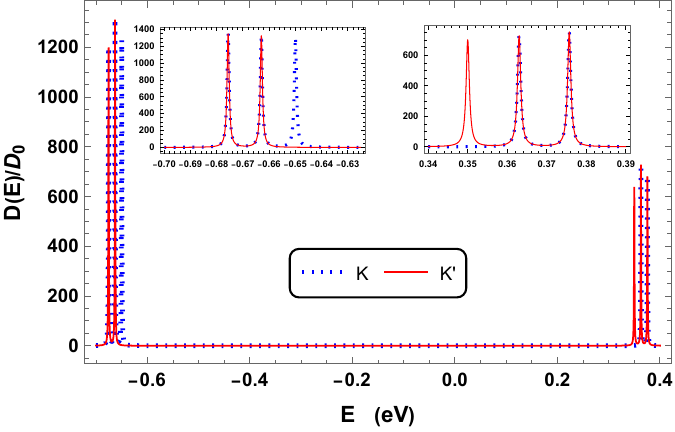}\\
    {\small\textbf{(a)}}  &  {\small\textbf{(b)}} \\[0.8em]
	\includegraphics[width=0.450\linewidth]{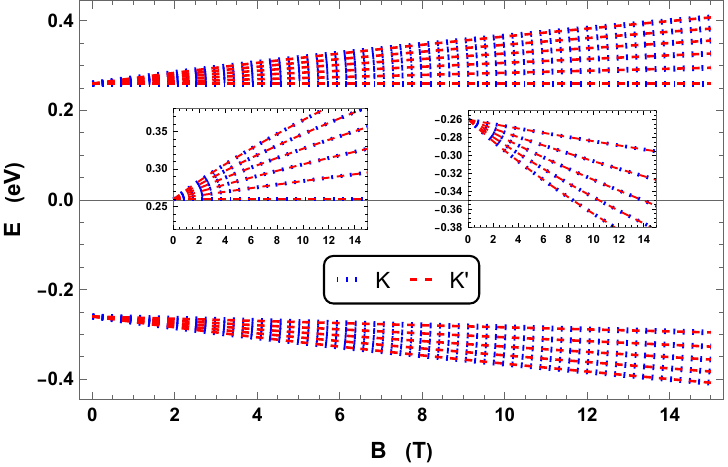}&
     \includegraphics[width=0.45\linewidth]{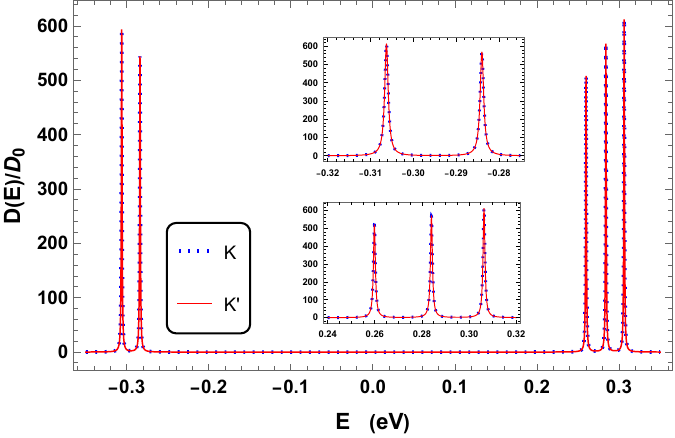}\\
     {\small\textbf{(c)}}  &  {\small\textbf{(d)}}  \\
    \end{tabular}}
	\caption{First column shows band structure of trivial BI  and TI phases of modified-Haldane model i.e. LL spectrum (in units of eV) as function of magnetic field B (in units of Tesla). Second column plots the DOS $(D(E)/D_0)$ for both phases. The upper(down) row shows trivial(topological) phase. The dotted blue (dashed red) curves show $K$ and $K'$ valleys. We have taken $B=10$ T and $\Gamma=0.5$ meV and $t_2=0.05$ eV. For trivial (topological) case, $M=0.05(0)$ eV.}
	\label{trivtopo}
\end{figure*}

We examine how the dispersion of the LLs varies with the parameters \( t_2 \), \( \phi \), and \( M \) by considering two representative cases. For simplicity, we fix NNN hopping strength to be $t_2=0.05$ eV in both cases. A trivial case with vanishing Haldane mass: \( \phi = 0 \) implying real NNN hopping and a topological case for Haldane mass dominating the Semenoff mass implying complex NNN hopping.

Trivial BI are characterized by a large intrinsic band gap originating from  sublattice inversion asymmetry $M$ together with real NNN hopping $t_2$. The energy dispersion relation simplifies to
\begin{align}
    E_{n,\eta}^{\tau} &= 
\begin{cases}
-3 t_2+\eta \sqrt{ 2v_F^2\hbar eB|n| + M^2}, & n \neq 0,\\
-3 t_2-\tau M, & n=0,
\end{cases} 
\end{align}
For $n\ge1$, the dispersion is independent of the valley index as shown in Fig. \ref{trivtopo}(a). 
In contrast, the LLL is valley-band polarized i.e. for the $ K(K')$ valley, the LLL lies in the VB (CB) as illustrated in the inset of Fig. \ref{trivtopo}(a). The fundamental band gap, i.e. the energy gap between the highest VB (LLL in $K$ valley) and the lowest CB (LLL in $K'$ valley) is given by $2M$. The dispersion is symmetric about $E=-3t_2$ rather than about $E=0$, since the real NNN interaction, $t_2$, breaks the particle-hole symmetry. 

By setting $\sin\phi \neq0$, the NNN hopping $t_2$ becomes complex. This breaks time-reversal symmetry and drives the system into a TI phase. To maximize topological effects, we take $\phi=\pi/2$ and keep inversion symmetry intact i.e. $M=0$, which will make LL spectrum valley independent as shown in Fig. \ref{trivtopo}(c). This is original Haldane model with only time-reversal breaking term leading to complex NNN hopping. The dispersion relation simplifies to
 \begin{align}
    E_{n,\eta} &= 
\begin{cases}
\eta \sqrt{ 2v_F^2\hbar eB|n| + (3\sqrt3 t_2\sin\phi)^2}, & n \neq 0,\\
3\sqrt3 t_2\sin\phi, & n=0.
\end{cases} 
\end{align}
All the LLs ($n\ge0$) are valley degenerate in contrast to the trivial case where the LLL has valley dependence. This implies the LLL in both valleys  lie in the CB as shown in inset of Fig. \ref{trivtopo}(c). In addition, the band gap is $6\sqrt3 t_2$ (symmetric about $ E=0$) with a small asymmetry of the order $\hbar v_F^2eB/3\sqrt3 t_2$. 
\begin{figure*}[ht!]
	\centering\scalebox{1}{
    \begin{tabular}{cc}
     \includegraphics[width=0.450\linewidth]{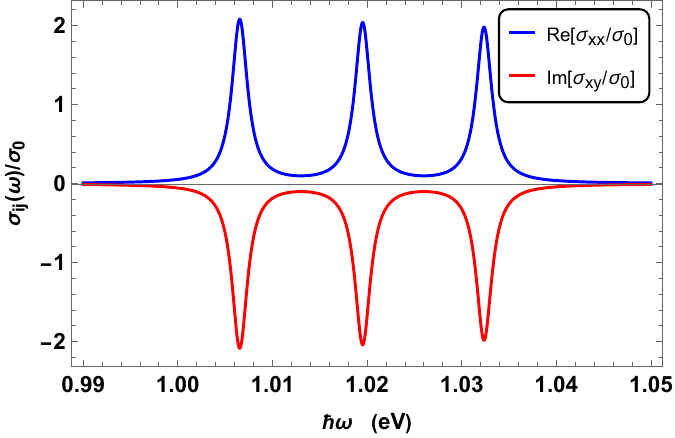}&
	\includegraphics[width=0.450\linewidth]{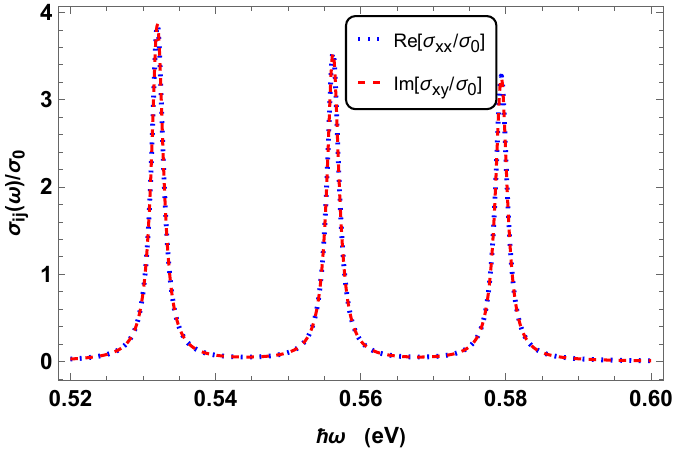}\\
    {\small\textbf{(a)}}  & {\small\textbf{(b)}} \\[0.8em]
    \includegraphics[width=0.45\linewidth]{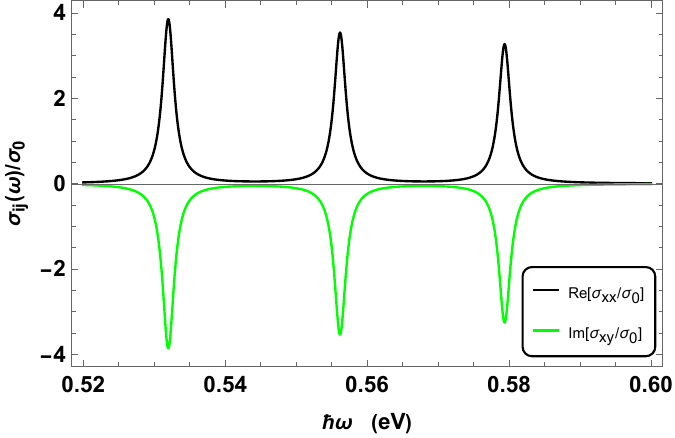}&
    \includegraphics[width=0.45\linewidth]{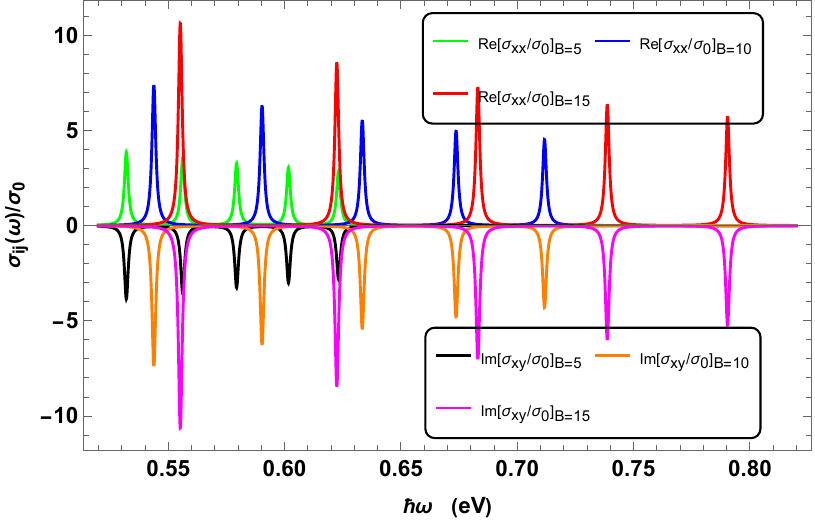}\\
    {\small\textbf{(c)}}  &  {\small\textbf{(d)}}
    \end{tabular}}
	\caption{(a) M-O conductivities for trivial phase. Both valleys show same M-O behavior. The blue  and red curves represent $\mathrm{Re[\sigma_{xx}/\sigma_0]}$ and $\mathrm{Im[\sigma_{xy}/\sigma_0]}$ respectively. The first peak corresponds to $\Delta_{0\rightarrow1}$ (b) and (c) show M-O conductivities for TI phase in $K$ and $K'$ valley respectively. Dotted blue (dashed red) represent $\mathrm{Re[\sigma_{xx}/\sigma_0]}$ and $\mathrm{Im[\sigma_{xy}/\sigma_0]}$  in $K$ valley and and black (green) in $K'$ valley. The first peak corresponds to $\Delta_{-1\rightarrow0}$. We have taken $B=5$ T and $\Gamma=1$ meV. (d) Effect of $B$ on M-O response. We have plotted conductivities for TI in $K'$ valley at 3 different values of $B$ i.e. 5 T, 10 T and 15 T).}
	\label{trivtopo1}
\end{figure*} 
\begin{figure}[ht!]
    \centering
    \begin{tabular}{c}
      \includegraphics[width=0.9\columnwidth]{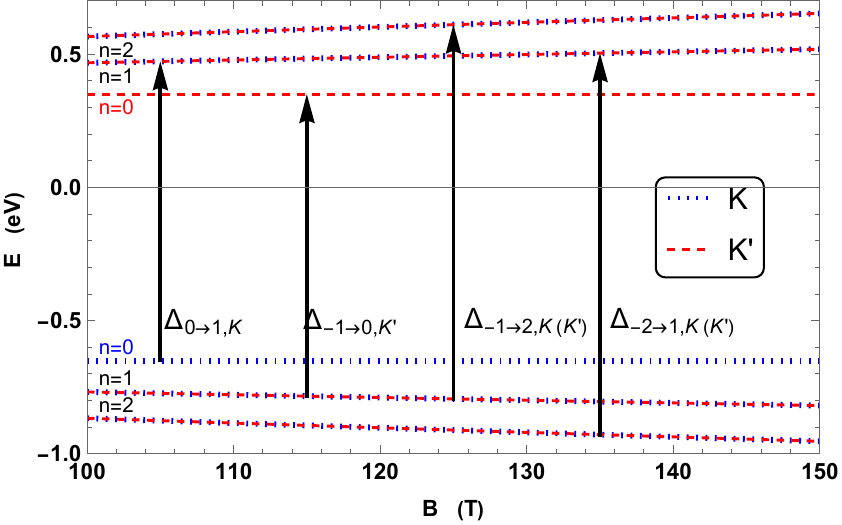} \\
    {\small\textbf{(a)}} \\[0.8em]
    \includegraphics[width=0.9\columnwidth]{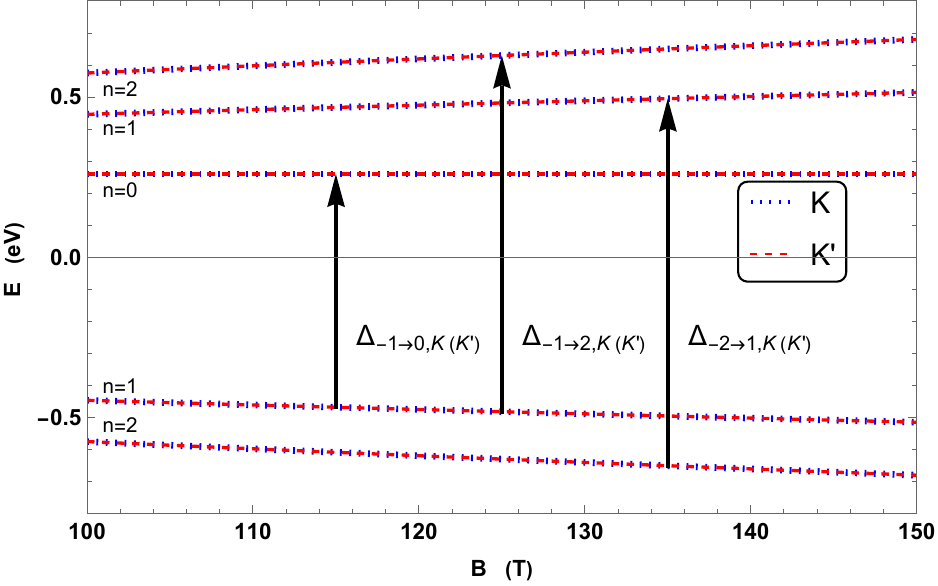} \\
    {\small\textbf{(b)}}
    \end{tabular}
   \caption{Allowed transitions for trivial BI and TI phases. (a) In trivial regime, LLL is located at VB (CB) in $K(K')$ valley. So $\Delta_{0\rightarrow1}$ and $\Delta_{-1\rightarrow0}$ both occur at energy . (b) In TI, LLL is at CB so only $\Delta_{-1\rightarrow0}$ is allowed in both valleys and $\Delta_{0\rightarrow1}$ is Pauli-blocked.}
    \label{trans}
\end{figure}
The valley-resolved DOS plotted in Fig. \ref{trivtopo}(b) for trivial case further confirms the fact that the LLL appears in different bands at two valleys, consistent with the dispersion shown in Fig. \ref{trivtopo}(a). The DOS exhibits symmetry (asymmetry) about $E=0$ in the topological(trivial) phase. The DOS is evaluated for Landau levels up to $n\leq2$ yielding three distinct peaks in each band in the trivial phase. Each valley chooses to place LLL in specific band. This behavior reflects the valley-dependent allocation of the LLL to a specific band. In the TI phase, however, one of the valence-band peaks (corresponding to LLL) disappears and is shifted to the CB, providing a clear signature of a topological phase transition. This redistribution indicates that the LLL is energetically favored in the CB in the TI regime. Such a shift in the Landau-level structure is expected to strongly influence the allowed M-O transitions, an effect that will be analyzed in detail later in this section. The corresponding valley-resolved DOS for topological case is shown in Fig. \ref{trivtopo}(d) where the third (left most in CB) peak shows presence of LLL in the CB for both valleys.

The contrasting energy dispersions of the BI and TI phases form the basis for understanding their M-O conductivities, which are intrinsically governed by their respective positions in the phase diagram (see Fig. \ref{phase}(c)). We restrict our discussion to absorptive part of M-O-response in this paper i.e. $\mathrm{Re[\sigma_{xx}]}$ and $\mathrm{Im[\sigma_{xy}]}$. These conductivities display a rich structure as functions of the photon frequency $\hbar\omega$ and feature multiple absorption resonances corresponding to allowed LL transitions for both spin channels. These peaks serve as fingerprints of the underlying LL structure, enabling extraction of band parameters and identification of the prevailing TI phase. The transition energies follow the electric-dipole selection rules $|n| - |m| = \pm 1$, together with spin conservation, such that inter-spin ($s=+1 \rightarrow s=-1$) transitions are strictly forbidden. Throughout this work, we focus on the undoped case ($\mu_F=0$), such that optical transitions are purely inter-band in nature. 

The absorption part of longitudinal and transverse M-O conductivities, i.e., the real part of $\mathrm{\sigma_{xx}}$ and the imaginary part of $\mathrm{\sigma_{xy}}$ are plotted in Fig. \ref{trivtopo1}(a). Both valleys show identical M-O response. Singular features in M-O conductivities arise when the incident photon energy satisfies the resonance condition $\hbar\omega = E_{n,\eta}^{\tau,s} - E_{m,\eta'}^{\tau,s}$, i.e., when the photon energy matches the magneto-excitation gap between the two LLs. The real (imaginary) component of $\mathrm{\sigma_{xx}}$ ($\mathrm{\sigma_{xy}}$) consists of a series of absorptive Lorentzian resonances whose full width at half maximum is dictated by the scattering rate $\Gamma$; increasing $\Gamma$ produces broader and reduced-amplitude peaks. Conversely, the real (imaginary) part of $\mathrm{\sigma_{xy}}$ ($\mathrm{\sigma_{xx}}$) exhibits dispersive Lorentzian line shapes. All resonances occur precisely at the magneto-excitation energies $\hbar\omega = E_{n,\eta}^{\tau,s} - E_{m,\eta'}^{\tau,s}$. However, for photon energies exceeding the band gap, the real part of $\mathrm{\sigma_{xx}}$ and the imaginary part of $\mathrm{\sigma_{xy}}$ decay monotonically, whereas in complementary components the logarithmic singularities persist. The corresponding M-O excitation energies associated with these transitions, given by \({E}^{\tau,s}_{n,\eta} - E^{\tau,s}_{m,\eta'} \), are labelled as \(\Delta^{\tau,s}_{m \rightarrow n} \) as shown in Fig. \ref{trans}(a). In this analysis, we focus exclusively on transitions between the $n=0,m=\pm1$ and $n=\pm1,m=\mp2$ LLs as the first ones (having $n=0$) solely depict the topology of the system and we have taken $\mu_F=0$. In the trivial phase, the LLL in the $K'(K)$ valley lies in CB (VB). So, the allowed transition in the $K(K')$ valley that includes LLL is $\Delta_{0\rightarrow1}(\Delta_{-1\rightarrow0})$; both having peaks at the excitation energy $\hbar\omega=1.00654$ eV. The corresponding energies for the lowest and first order allowed transitions are tabulated in Table \ref{alltrans}.
\begin{table}[h]
    \centering
    \caption{Table of M-O excitation energies for allowed transitions in the $n=(-2,-1,0,1,2)$ subspace for $B=5$ T in trivial BI and TI phases as shown in Fig. \ref{trivtopo1}(a,b,c)}
    \label{trans1}
    \begin{tabular}{ccc}
        \toprule
        Regime& $\Delta_{m \rightarrow n}$
& Photonic energy (eV)\\
        \midrule
        Trivial & $\Delta_{ 0\rightarrow1,K}=\Delta_{ -1\rightarrow0,K'}$& 1.00654\\
        Trivial & $\Delta_{-1\rightarrow2,K(K')}$& 1.01953\\
        Topological & $\Delta_{-1\rightarrow0,K(K')}$& 0.531984\\
        Topological & $\Delta_{-1\rightarrow2,K(K')}$& 0.556183\\
        \bottomrule
    \end{tabular}
    \label{alltrans}
\end{table}

Fig. \ref{trivtopo1}(b) and (c) displays the absorptive peaks i.e. the real and imaginary components of $\mathrm{\sigma_{xx}}$ and $\mathrm{\sigma_{xy}}$ for the topological case, as a function of photon energy,  $\hbar\omega$, in $K$ and $K'$ valley respectively. In the TI, LLL in both valleys lies in CB, so only $\Delta_{-1\rightarrow0}$ is allowed in both valleys at the energy $\hbar\omega=0.531984$ eV. The prohibition of the  optical transition $\Delta_{0\rightarrow1}$ in the $K$ valley is a direct consequence of the underlying topology of the system and is insensitive to material choice. The corresponding excitation energies for allowed transitions are summarized in Table \ref{alltrans}. In each of the depicted transitions, the transition that distinguishes the topology of the material is the one in which one of the participating levels is LLL. The transition $\Delta_{-1\rightarrow2}$ or $\Delta_{-2\rightarrow1}$ and higher LL transitions are always present in all materials, although their spectral weights vary.

To illustrate the influence of the magnetic field on the energy-level structure and the resulting M-O response, schematic diagram of the allowed transitions between LLs in the $K'$ valley, corresponding to magnetic fields $B=5,10 \text{ and } 15$ T (all within the TI regime), are presented in Fig. \ref{trivtopo1}(d). As the magnetic field strength increases, the associated excitation energies shift toward higher photon energies, leading to displacement of the peaks in the conductivity spectra. Furthermore, due to the redistribution of spectral weight among the transitions, the peak intensities increase with higher magnetic fields. To summarize;  in the TI phase, energy dispersion is valley-degenerate whereas the M-O-response is valley dependent. In contrast, in the trivial phase the energy dispersion becomes valley dependent while the M-O-response becomes valley independent.
 
  Now we proceed to show that the modified-Haldane model, studied above, is a general model that encompasses effective models for various 2D Dirac materials. One can obtain the M-O response of these materials by tuning parameters in the modified-Haldane model.\\
\subsection{Topological silicene} 
Buckled Xene monolayers \cite{doi:10.7566/JPSJ.84.121003,buckled} (X refers to group-IV elements) including silicene, germanene, stanene, and plumbene, are graphene analogs composed of silicon, germanium, tin, and lead, respectively. Unlike graphene, these materials have a buckled out-of-plane lattice structure, which gives rise to a significantly enhanced intrinsic SOC. The SOC strength $\Delta_{so}$ is approximately 1.55-7.9 meV in silicene\cite{ezawa2012spin}, 24-93 meV in germanene\cite{PhysRevLett.107.076802} and 100 meV in stanene \cite{xu2013large}.
An electric field perpendicular to the buckled Xene plane breaks inversion symmetry and induces a staggered sublattice potential $M$ between the  $A$ and $B$ sublattices. Consequently, the interaction between the $M$ and buckled lattice renders the Dirac fermion mass tunable at both valleys, enabling multiple topological phase transitions \cite{duan2018bulk,Khan:25,PhysRevB.101.205408}. By substituting $t_2 = \Delta_{so}/3 \sqrt 3\text{ and } \phi= \pm\pi/2$ into the dispersion relation of the generic Hamiltonian Eq.\eqref{energy}, one recovers the low-energy dispersion of the buckled Xene monolayers as \cite{tabert2013magneto,tahir2015magneto,Khan:25,ezawa2012spin,Ezawa_2012,Ezawa2012a,Ezawa2012,ezawa2012topological}
\\
\begin{align}
    E_{n,\eta}^{\tau,s} &= 
\begin{cases}
\eta \sqrt{ 2v_F^2\hbar eB|n| + (M-\tau s \Delta_{so})^2}, & n \neq 0,\\
-\tau(M-\tau s \Delta_{so}), & n=0.
\end{cases} 
\end{align}
Here $\Delta_{so}$ denotes intrinsic SOC while $M$ represents the electrically tunable band gap. The relative magnitude of $M$ with respect to $\Delta_{so}$  determines the TI phase of the system. The phases $\phi = \pi/2$ and $\phi = -\pi/2$ correspond to spin-up($s=1$) and spin-down($s=-1$) states respectively. The Rashba SOC and Zeeman terms are typically an order of magnitude smaller than the intrinsic SOC and can therefore be neglected to the leading order. Consequently, this dispersion provides a well-approximated low-energy physics of silicene.

The intrinsic SOC is responsible for opening a gap $2\Delta_{so}$ between spin-resolved bands, irrespective of the magnitude of $M$, which introduces a valley-asymmetric mass term. The competition between these two energy scales determines the TI phase of the system. When $M < \Delta_{so}$, the effective Dirac masses associated with same spin-states acquire same signs, leading to a band inversion at the Dirac points for certain spin-valley sectors and the realization of a TI phase. In contrast, when $M>\Delta_{so}$, the mass terms share the same sign across valleys, suppressing band inversion resulting in a trivial BI phase.

 For the case $M=0$, the energy bands remain valley degenerate. LLLs are spin-split by $\Delta_{so}$ while higher LLs acquire spin degeneracy in addition to valley degeneracy. When $M$ is introduced, the LLLs become spin-valley selective: the spin-down(up) state in $K(K')$ valley resides in VB(CB) irrespective of the relative magnitudes of $M$ and $\Delta_{so}$. The Kramers pair with $s\tau=1$ i.e. $(K,\uparrow)$ and $(K',\downarrow)$ is sensitive to competition between $M$ and $\Delta_{so}$, and its band location determines the TI phase. This differentiating pair switches between the valence and conduction bands as $M$ crosses $\Delta_ {so}$, while its complementary pair with $s\tau=-1$ remains fixed in band although gap between them increases, highlighting the role of the former as a topological indicator. The gap associated with this pair (its counterpart) is symmetric about zero energy and equals $2|\Delta_{so} - M|( 2|\Delta_ {so} + M|)$.

 When $M<\Delta_{so}$, $(K,\uparrow)$ and $(K',\downarrow)$ are in CB and VB respectively. This implies spin-up LLLs lie in VB while spin-down LLLs lie in CB making bands spin-selective. The system is said to be in QSHI phase \cite{ezawa2012spin,Khan:25,PhysRevLett.107.076802,tahir2015magneto,Wang_2014}. As $M$ increases, the topologically important pair gets closer to each other, while other pair gets further apart. At $M=\Delta_{so}$, the topologically important pair lies exactly at zero energy, and this TI phase is termed VSPM \cite{Ezawa2012,tahir2015magneto}. Increasing M further makes $(K,\uparrow)$ reside in VB and $(K',\downarrow)$ reside in CB, therefore, the bands are now valley selective for LLL. The gaps reopen without band inversion, and the system transitions to a trivial (BI) phase. Thus, buckled Xene monolayers exhibit a highly tunable topological phase diagram controlled by the $M$ and intrinsic SOC.

\subsubsection{Topological Phases}
\paragraph{\textbf{Quantum spin Hall insulator (QSHI) : }}
In silicene, the LL spectrum is controlled by the intrinsic SOC $\Delta_{so} $ and the buckled lattice geometry parametrized by $M$. Tuning the parameter $M$ so that it equals $\Delta_{so}/2$, makes the system QSHI \cite{PhysRevLett.107.076802}, which couples spin ( $s=\pm1$ ) and valley ($ \tau=\pm1$ ) through an effective mass  \cite{Wang_2014,Khan:25,ezawa2012spin,Khan:25,PhysRevLett.107.076802,tahir2015magneto}. The resulting LL spectrum is
 \begin{align}
    E_{n,\eta}^{\tau,s} &= 
\begin{cases}
\eta \sqrt{ 2v_F^2\hbar eB|n| + \Delta_{so}^2(1/2-\tau s)^2}, & n \neq 0,\\
 (s-\tau/2) \Delta_{so}, & n=0,
\end{cases} 
\end{align}
where ( $\eta=\pm1$ ) labels the conduction/valence branches.
\begin{figure}[ht!]
        \centering
    \begin{tabular}{c}
      \includegraphics[width=0.9\columnwidth]{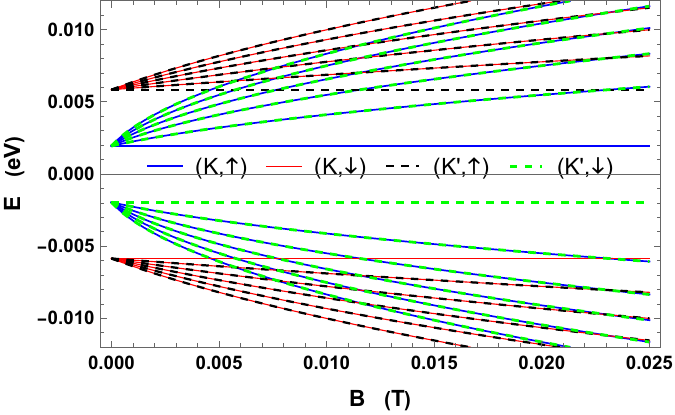} \\
    {\small\textbf{(a)}} \\[0.8em]
    \includegraphics[width=0.9\columnwidth]{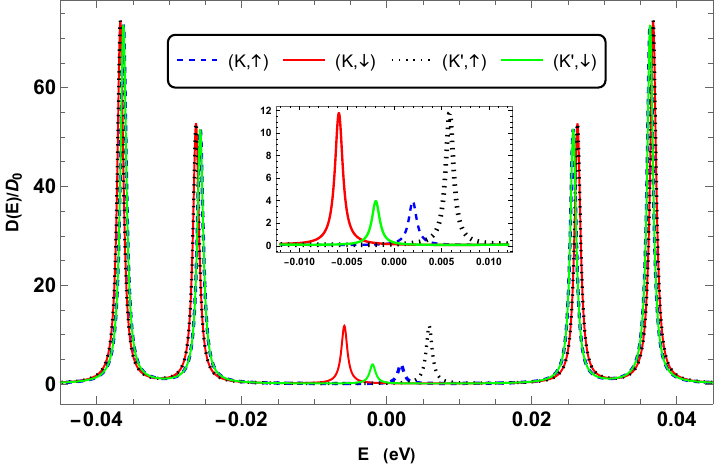} \\
    {\small\textbf{(b)}}
    \end{tabular}
   \caption{(a) LL spectrum of silicene in QSHI phase (in units of eV) as function of magnetic field B (in units of Tesla) parametrized by $M = 0.5\Delta_{so}$ at the $K$ and $K'$ valleys with up and down spin energy bands shown simultaneously. (b) The DOS  $(D(E)/D_0)$ at two in-equivalent valleys $K$ and $K'$ shows negligible spin-splitting within each valley. The blue(red) curves represent up(down) spins in $K$ valley and black(green) curves represent up(down) spins in $K'$ valley. We have taken $B=0.5$ T and $\Gamma=0.5$ meV.}
	\label{qshi}
\end{figure}

The LL spectrum is plotted in Fig. \ref{qshi}(a) showing spin, valley, and band dependence. The LLL exhibits a characteristic spin-valley dependence arising from intrinsic SOC and the buckled lattice geometry. Since  $(s - \tau/2)\Delta_{so} $ changes sign with both spin and valley, the LLL becomes strongly spin-valley polarized. Specifically, the CB hosts the spin-up states from both $K$ and $K'$ valleys, while the valence band (VB) accommodates the corresponding spin-down states from both valleys. This asymmetric placement of the LLL reflects the opposite effective masses generated for the two spin species. The LLL inherits this mass asymmetry and becomes both spin and valley-selective. This spin-valley locking of the LLL is a defining feature of the QSHI phase and directly reflects the opposite signs of the SOC-induced Dirac masses at $K$ and $K'$.

For $n\ge1$, the higher LLs appear symmetrically in both the VB and the CB for each $\tau$ and $s$ and Kramer's doublet becomes spin degenerate. However, unlike LLL, the SOC-induced splitting between both pairs vanishes for $n\ge1$ as both bands host these polarized states, which can be seen in Fig. \ref{qshi}(a).

Consequently, the QSHI phase of silicene is characterized by a fully spin-valley-resolved LLL confined to a specific band depending on spin, while the higher LLs form quasi-degenerate ladders of Kramer's pair in both bands. This distinct separation between the LLL and the higher LLs is a central signature of SOC-driven TI phases in buckled honeycomb materials.

The electronic DOS for 2D hexagonal structures in the $K$ and $K'$ valleys across different TI phases has also been plotted. Fig.~\ref{qshi}(b) shows the overall DOS \cite{Khan:25,tabert2013magneto} in the in-equivalent valleys $K$ and $K'$  for the spin-up and spin-down states and the inset shows the DOS of LLL.

The DOS is symmetric about Fermi energy. Four spin- and valley-polarized peaks emerge at the energies corresponding to the LLLs. The topologically important pair exhibits shorter DOS peaks than their counterparts due to the smaller Dirac mass (band-gap).

The higher LLs exhibit an approximate degeneracy among Kramer's pairs, and have taller peaks indicating a higher degeneracy than LLLs. The higher LLs draw weights symmetrically in both bands and give a larger effective contribution to the DOS due to larger mass. The peak-height ordering thus provides a direct spectroscopic signature of spin-valley mass inversion in the QSHI phase.

\begin{figure}[ht!]
    \centering
        \begin{tabular}{c}
\includegraphics[width=0.95\columnwidth]{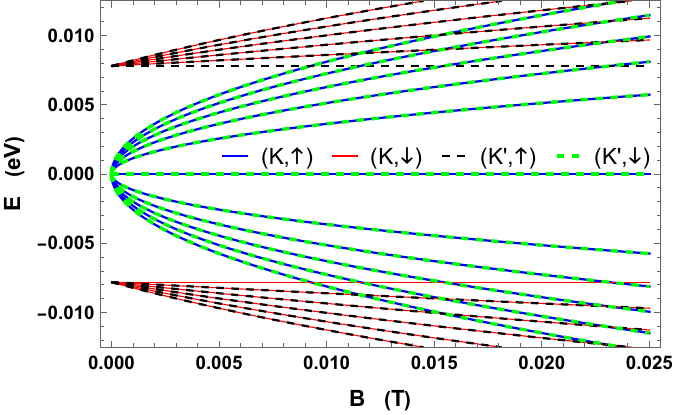}\\
        {\small\textbf{(a)}} \\[0.8em]
		 \includegraphics[width=0.95\columnwidth]{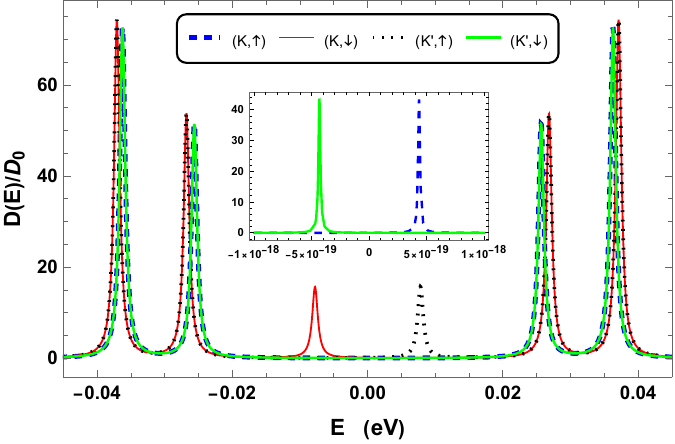}\\
           {\small\textbf{(b)}}
           \end{tabular}
    \caption{(a) LL spectrum of silicene in valley-spin polarized metal (VSPM) phase as function of magnetic field $B$ parametrized by $M = \Delta_{so}$. The spectra at the $K$ and $K'$ valleys are shown simultaneously, including both spin-up and spin-down bands. (b) The overall DOS $(D(E)/D_0)$  at two in-equivalent valleys $K$ and $K'$ is shown. The blue(red) curves represent up(down) spins in $K$ valley and black(green) curves represent up(down) spins in $K'$ valley. We have taken $B=0.5$ T and $\Gamma=0.5$ meV. However, the peaks in the inset are only resolved at $\Gamma=10^{-20}$ eV.}
    \label{vspm}
\end{figure}

\paragraph{\textbf{Valley spin polarized metal (VSPM) : }}
When the two competing energy scales are equal, the resulting LL spectrum transforms to \cite{Ezawa2012,tahir2015magneto}
 \begin{align}
    E_{n,\eta}^{\tau,s} &= 
\begin{cases}
\eta \sqrt{ 2v_F^2\hbar eB|n| + \Delta_{so}^2(1-\tau s)^2}, & n \neq 0,\\
 (s-\tau) \Delta_{so}, & n=0,
\end{cases} 
\end{align}
where $s=\pm1$ is the spin index and ($\tau=\pm1$) labels the valleys $K$ and $K'$ and is shown in Fig. \ref{vspm}(a).
The LLL structure is fully determined by the mass term $(s-\tau)\Delta_{so}$, producing strong spin-valley polarization, as shown in Fig. \ref{vspm} (a).
The gap between topologically important pair vanishes and it becomes degenerate exactly at zero energy which is also fixed as Fermi level. The gap between other pair increases to $4\Delta_{so}$.

This key feature  gives this phase the name of valley-spin-polarized metal in which one spin species from each valley lies exactly at the Fermi level $\mu_F=0$. These zero-energy states demonstrate explicit spin-valley locking: spin-up zero-energy state resides in $K$ valley while spin-down zero-energy state resides in $K'$. Higher LLs behave similarly as in QSHI except the band gap goes to zero as $B\rightarrow0$. 
This interplay of SOC, inversion asymmetry, and valley-contrasting mass terms produces a LL structure in which the system is insulating for one set of spin-valley polarization, but maintains symmetry-protected spin-valley polarized zero energy modes at the Fermi level, a hallmark of TI phases in buckled honeycomb lattices.
\begin{figure}[ht!]
        \centering
    \begin{tabular}{c}
      \includegraphics[width=0.9\columnwidth]{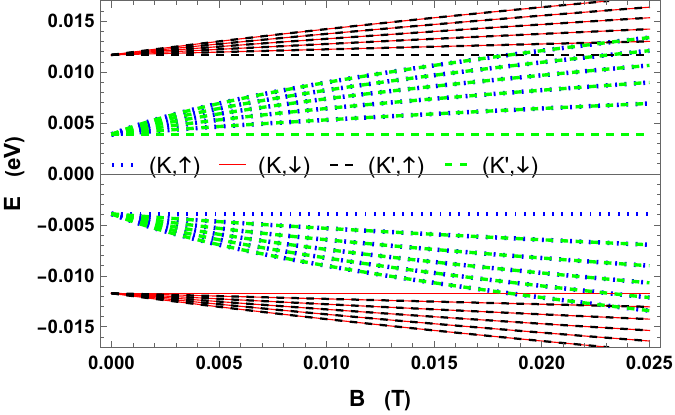} \\
    {\small\textbf{(a)}} \\[0.8em]
    \includegraphics[width=0.9\columnwidth]{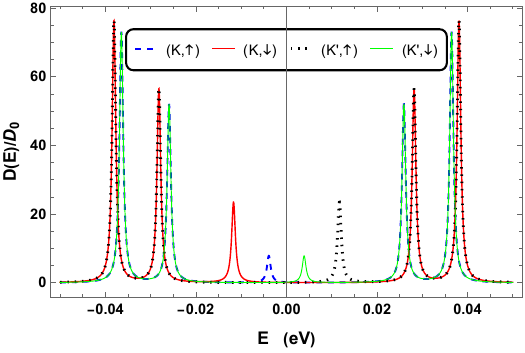} \\
    {\small\textbf{(b)}}
    \end{tabular}
    \caption{(a) Band structure of silicene in  band insulator (BI) phase parametrized by $M = 2\Delta_{so}$. LLL shows no band inversion. (b) The total DOS show four LLLs and the higher peaks for higher LLs.  The blue (red) curves represent up (down ) spins in $K$ valley and black (green) curves represent up (down) spins in $K'$ valley. We have taken $B=0.5$ T and $\Gamma=0.5$ meV.}
    \label{bi}
\end{figure}
\\
Fig.~\ref{vspm}(b) displays the valley- and spin-resolved DOS in the VSPM phase as a function of photon energy for both valleys. In the VSPM state, the two central LLL peaks associated with $(K',\uparrow)$ and $(K,\downarrow)$ appear as the dominant peaks in the DOS. The remaining peaks corresponding to topologically important pair i.e., $(K,\uparrow)$ and $(K',\downarrow)$ persist as an extremely weak feature and are visible only under ultra-small level broadening $\Gamma\sim10^{-20}$ eV, and even then their peak height is strongly suppressed due to the disappearance of the Dirac mass in these channels. This strong asymmetry between the LLLs peaks is a characteristic signature of the VSPM point, where half of the spin-valley sectors become massless while the others retain a finite mass.\\
\paragraph{\textbf{Band insulator (BI) :}}
When the staggered sub-lattice potential $M$ exceeds the intrinsic SOC scale $\Delta_{so}$ i.e. $M=2\Delta_{so}$. the dispersion relation modifies as
\begin{align}
    E_{n,\eta}^{\tau,s} &= 
\begin{cases}
\eta \sqrt{ 2v_F^2\hbar eB|n| + \Delta_{so}^2(2-\tau s)^2}, & n \neq 0,\\
 (s-2\tau) \Delta_{so}, & n=0. 
\end{cases} 
\end{align}
In this insulating regime, the effective Dirac mass for each spin-valley flavor is given by $\Delta_{\tau,s}=M-\tau s\Delta_{so}$, which gives $\Delta_{\tau,s}\ge0$ for all combinations of $\tau,s$. The corresponding LL energies reduce to $(s-2\tau) \Delta_{so}$, leading to LLL at $K(K')$ valley residing in CB(VB).

 This is because, the magneto-excitation energies for both spin channels are negative in the $K$ valley, whereas they lie at positive energy in the $K'$ valley, as depicted in Fig.~\ref{bi}(b). This arrangement signifies the absence of band inversion, reflecting the restoration of a trivial BI phase characterized by a fully opened gap and conventional spin-valley polarization of the LLs.
 The SOC-induced spin splitting survives but does not induce spin-valley locking or topological protection, in contrast to the $M <\Delta_{so}$ regime characteristic of the Kane-Mele TI phase. 

\subsubsection{Berry Curvature}
\begin{figure}[ht!]
    \centering
    \includegraphics[width=0.95\columnwidth]{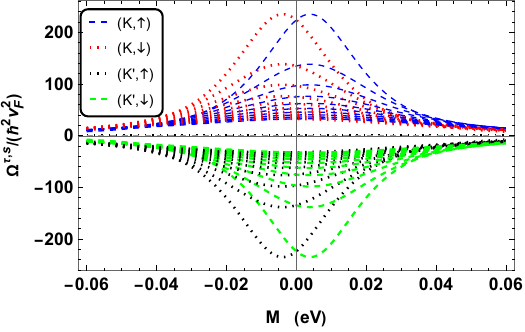}
    \caption{Berry Curvature $\Omega^{\tau,s}$ of silicene against $M$. The blue(red) curves represent $\Omega$ for up(down) spins in $K$ valley while black(green) curves represent $\Omega$ for up(down) spins in $K'$ valley. }
    \label{bcsili}
\end{figure}

Before proceeding towards M-O conductivities, let's discuss another important quantity; Berry curvature of silicene as a function of $M$. The expression for Berry curvature $\Omega_{n}^{\tau,s}$ is given in Eq.\eqref{a72} revealing strict valley-dependence. The Berry curvature exhibits a coupled spin-valley character and is strongly enhanced as $M$ approaches zero. $\Omega_{n}^{\tau,s}$ vanishes when inversion and time reversal symmetries survive.
 Symmetry arguments lead to
\begin{equation}\label{a73}
\Omega^{K,\uparrow}=-\Omega^{K',\downarrow}, \quad \Omega^{K,\downarrow}=-\Omega^{K',\uparrow}.
\end{equation}
Eq.\eqref{a72} tells that the Berry curvature $\Omega_{n}^{\tau,s}$ can be tuned by an external field. Fig. \ref{bcsili} shows the Berry curvatures of silicene as a function of $M$, revealing distinct responses for different spin-valley sectors. $\Omega_{n}^{\tau,s}$ is positive in the $K$ valley and negative in the $K'$ valley and follow relation of Eq. \eqref{a73}, as illustrated in Fig.~\ref{bcsili}.

\begin{figure*}[ht!]
    \centering\scalebox{1}{
    \begin{tabular}{cc}
    \includegraphics[width=0.45\linewidth]{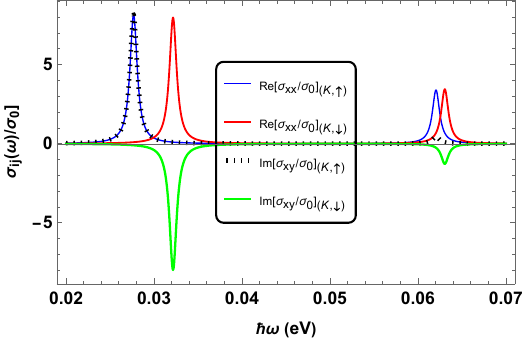}&
    \includegraphics[width=0.45\linewidth]{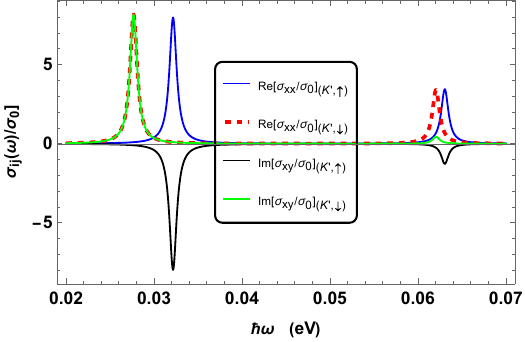}\\
       {\small\textbf{(a)}}  &  {\small\textbf{(b)}}  \\[0.8em]
    \includegraphics[width=0.45\linewidth]{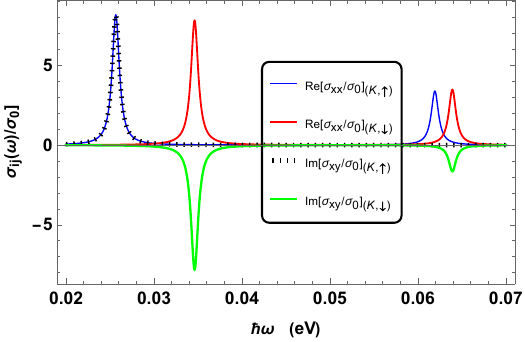}&
    \includegraphics[width=0.45\linewidth]{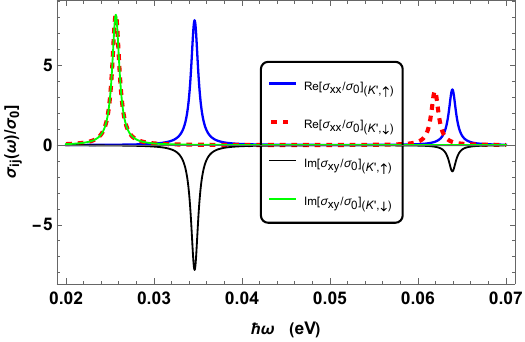}\\
       {\small\textbf{(c)}}  &  {\small\textbf{(d)}}  \\[0.8em]
    \includegraphics[width=0.45\linewidth]{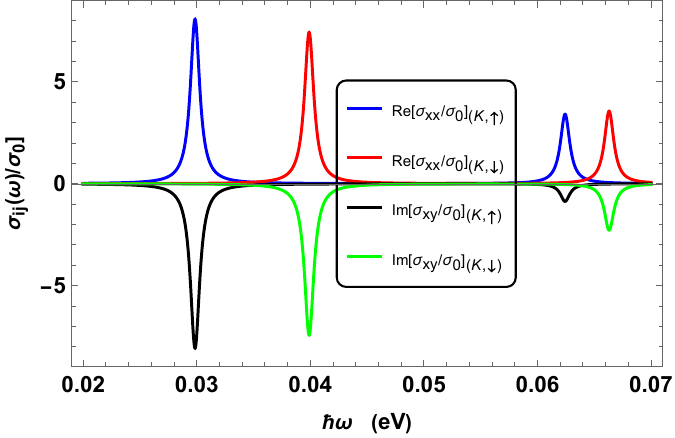}&
    \includegraphics[width=0.45\linewidth]{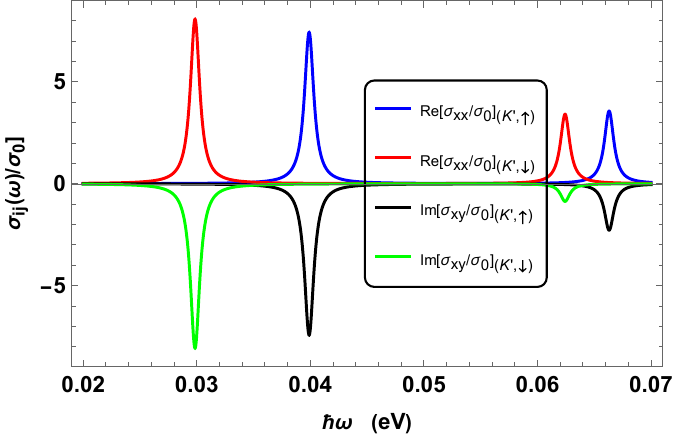}\\
       {\small\textbf{(e)}}  &  {\small\textbf{(f)}}  \\
    \end{tabular}
    }
    \caption{M-O conductivities $\mathrm{Re[\sigma_{xx}/\sigma_0]}$ and $\mathrm{Im[\sigma_{xy}/\sigma_0]}$ of different phases of silicene depending upon values of $M$ relative to $\Delta_{so}$. (a,b) M-O conductivities of QSHI phase in $K$ and 
$K'$ valley. (c,d) M-O conductivities of VSPM in $K$ and $K'$ valley. (e,f) M-O conductivities of BI in $K$ and $K'$ valley. Left(right) panel shows response of $K(K')$ valley. The blue(red) curves represent $\mathrm{Re[\sigma_{xx}/\sigma_0]}$ for up(down) spins and black(green) curves represent $\mathrm{Im[\sigma_{xy}/\sigma_0]}$ for up(down) spins. The first peaks correspond to transitions involving LL and determine the topology of phase, while second peaks correspond to higher level transitions. We have taken $B=0.5$ T and $\Gamma=0.5$ meV. }
        \label{sigmasilicene}
\end{figure*}

\begin{figure*}[ht!]
    \centering\scalebox{1}{
    \begin{tabular}{cc}
    \includegraphics[width=0.45\linewidth]{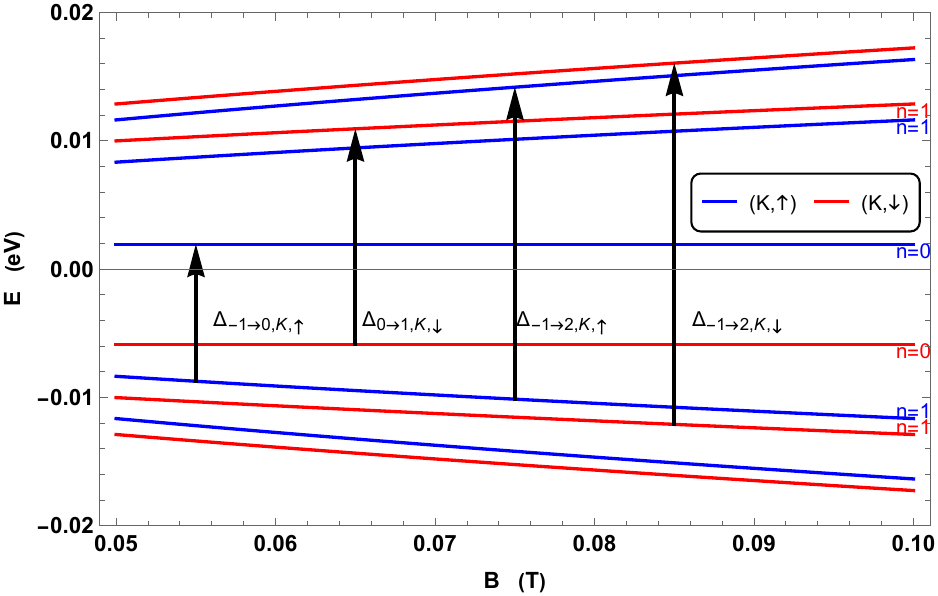}&
    \includegraphics[width=0.45\linewidth]{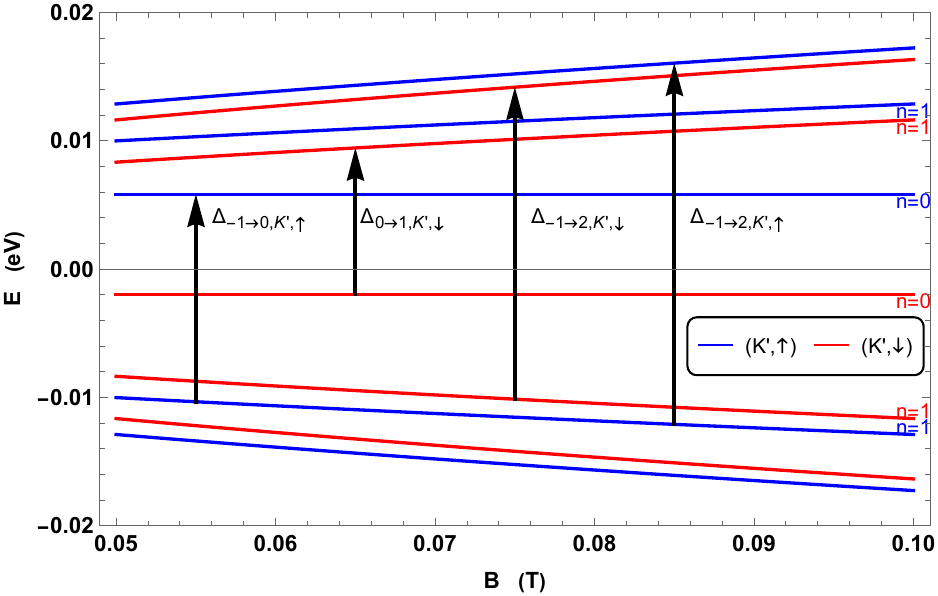}\\
    {\small\textbf{(a)}} & {\small\textbf{(b)}}\\[0.8em]
    \includegraphics[width=0.45\linewidth]{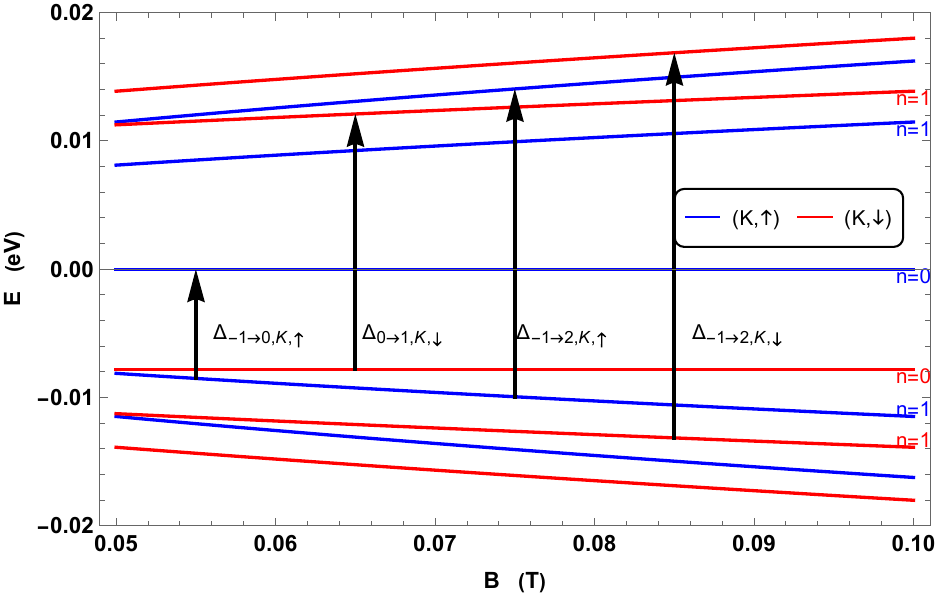}&
    \includegraphics[width=0.45\linewidth]{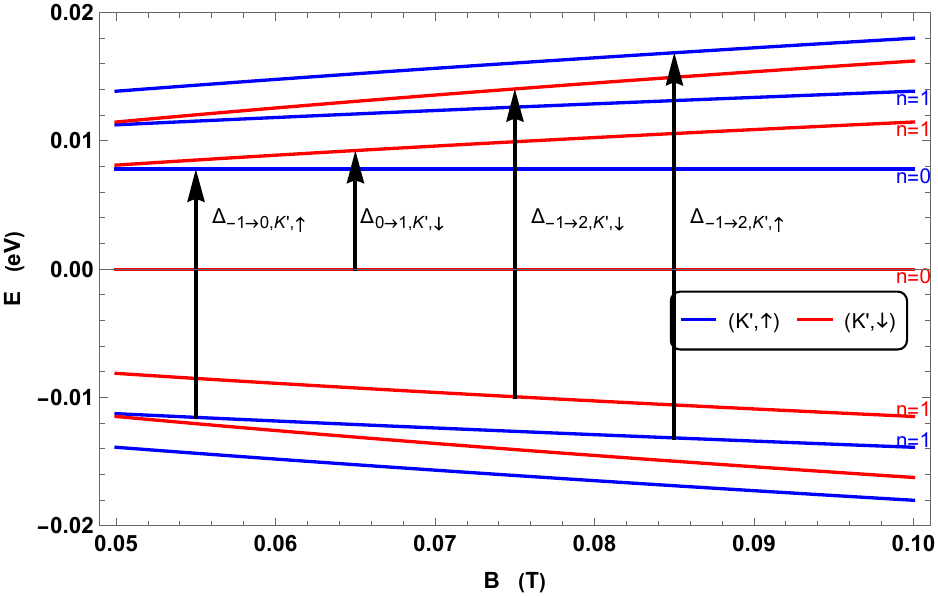}\\
    {\small\textbf{(c)}} & {\small\textbf{(d)}}\\[0.8em]
     \includegraphics[width=0.45\linewidth]{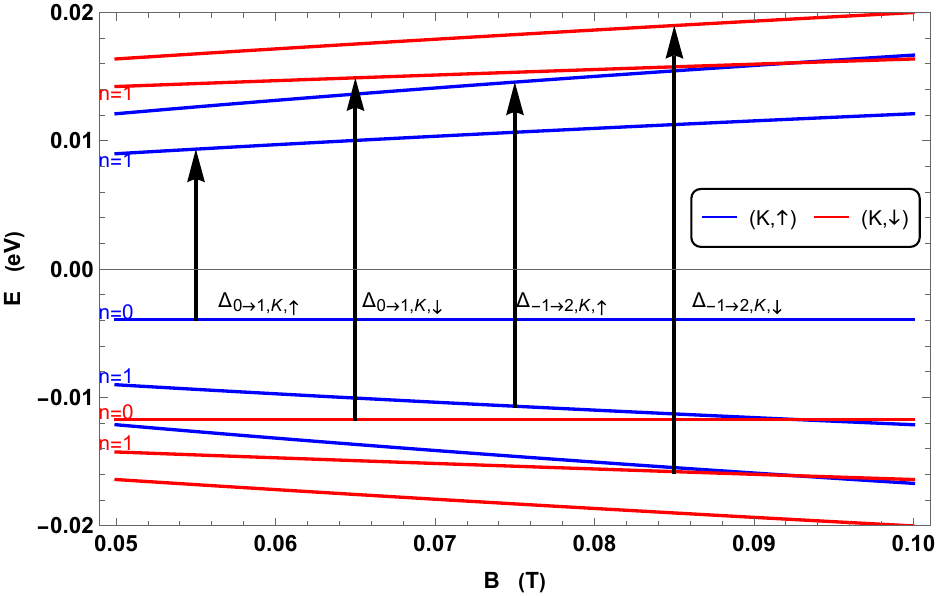}&
    \includegraphics[width=0.45\linewidth]{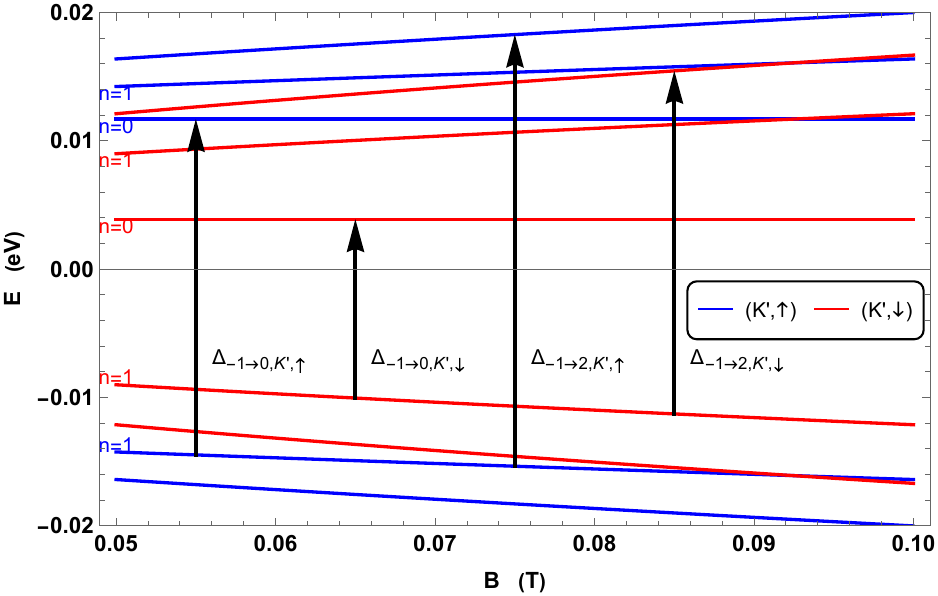}\\
    {\small\textbf{(e)}} & {\small\textbf{(f)}}\\
    \end{tabular}
    }
    \caption{Different rows exhibit allowed transitions in QSHI, VSPM and BI phases respectively. The left(right) panel shows the transitions in $K(K')$ valley. The blue(red) lines represent up(down)-spin states in both CB and VB.}
    \label{transsili}
\end{figure*}

\begin{figure*}[ht!]
    \centering\scalebox{1}{
    \begin{tabular}{cc}
      \includegraphics[width=0.450\linewidth]{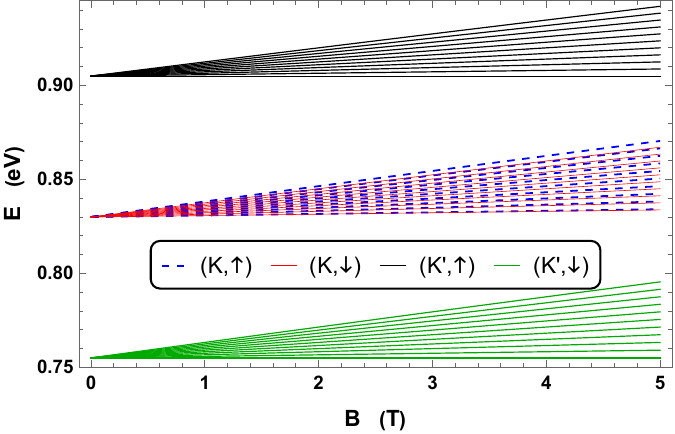}&
    \includegraphics[width=0.450\linewidth]{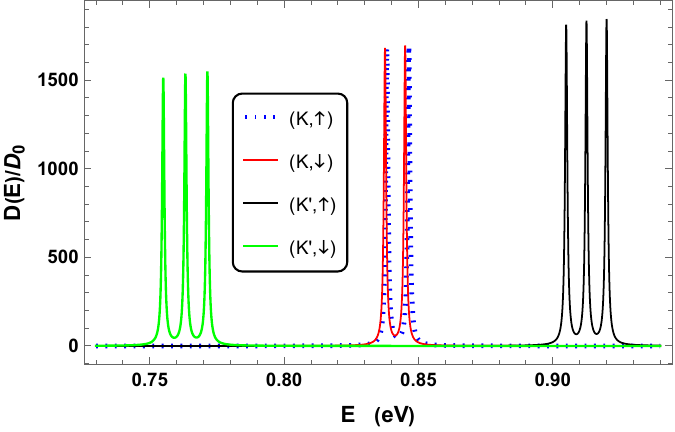}\\
    {\small\textbf{(a)}} & {\small\textbf{(c)}}\\[0.8em]
    \includegraphics[width=0.45\linewidth]{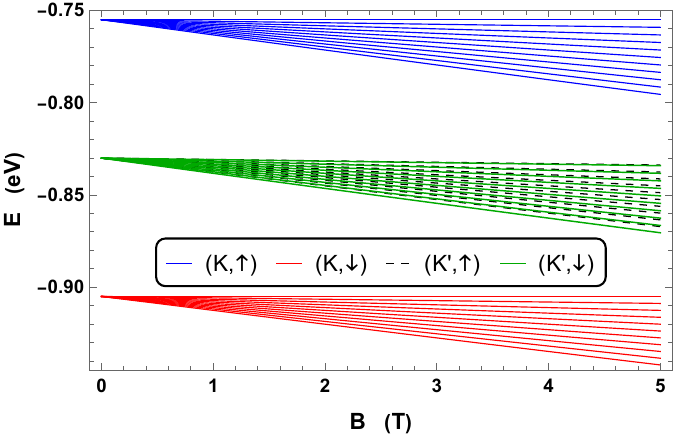}&
    \includegraphics[width=0.45\linewidth]{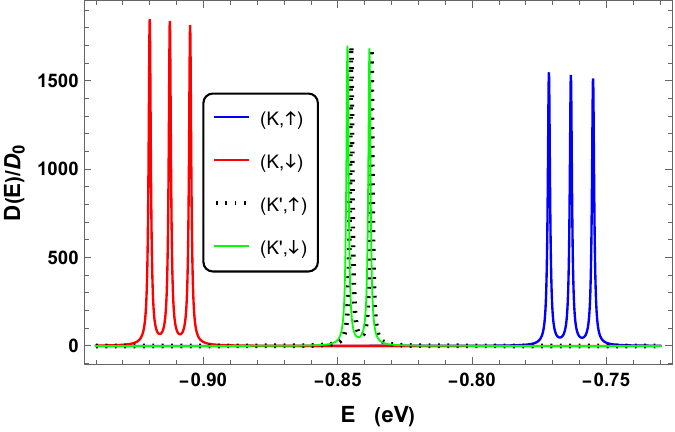}\\
    {\small\textbf{(b)}}& {\small\textbf{(d)}} \\
    \end{tabular}}
        \caption{(a,b) The LL spectrum of TMDC as function of  B is shown in CB and VB. The bands are drawn separately to clearly resolve the splittings as band gap is huge. LLL is shifted form VB to CB as we shift from $K$ to $K'$ valley. (c,d) The DOS $(D(E)/D_0)$ for both bands are plotted. The blue(red) curves represent up(down ) spins in $K$ valley and black(green) curves represent up(down ) spins in $K'$ valley. We have taken $\Delta_{tmdc}=75$ meV, $\Delta=1.6$ eV, $B=10$ T and $\Gamma=0.5$ meV.}
    \label{tmdc}
\end{figure*}

\begin{table}[h]
    \centering
    \caption{Table of M-O excitation energies for allowed transitions in the $n=(-2,-1,0,1,2)$ subspace for $B=0.5$ T in QSHI as shown in Fig. \ref{sigmasilicene}(a,b) and Fig. \ref{transsili}(a,b).}
    \label{trans3}
    \begin{tabular}{ccc}
        \toprule
         $\Delta_{m\rightarrow n, K,\uparrow(\downarrow)}$& $\Delta_{m\rightarrow n, K',\uparrow(\downarrow)}$& Photonic energy (eV)\\
        \midrule
         $\Delta_{-1\rightarrow0, K,\uparrow}$& $\Delta_{0\rightarrow1, K',\downarrow}$& 0.0276755\\
         $\Delta_{0\rightarrow1, K,\downarrow}$& $\Delta_{-1\rightarrow0, K',\uparrow}$& 0.0321601\\
         $\Delta_{-1\rightarrow2, K,\uparrow}$& $\Delta_{-1\rightarrow2, K',\downarrow}$& 0.0620546\\
         $\Delta_{-1\rightarrow2, K,\downarrow}$& $\Delta_{-1\rightarrow2, K',\uparrow}$
& 0.0630555\\
        \bottomrule
    \end{tabular}
    \label{qshitrrans}
\end{table}

\begin{table}[h]
    \centering
    \caption{Table of M-O excitation energies for allowed transitions  in the $n=(-2,-1,0,1,2)$ subspace for $B=0.5$ T in VSPM as shown in Fig. \ref{sigmasilicene}(c,d) and Fig. \ref{transsili}(c,d).}
    \label{trans4}
    \begin{tabular}{ccc}
        \toprule
         $\Delta_{m\rightarrow n, K,\uparrow(\downarrow)}$& $\Delta_{m\rightarrow n, K',\uparrow(\downarrow)}$& Photonic energy (eV)\\
        \midrule
         $\Delta_{-1\rightarrow0, K,\uparrow}$& $\Delta_{0\rightarrow1, K',\downarrow}$& 0.0256515\\
         $\Delta_{0\rightarrow1, K,\downarrow}$& $\Delta_{-1\rightarrow0, K',\uparrow}$& 0.0346112\\
         $\Delta_{-1\rightarrow2, K,\uparrow}$& $\Delta_{-1\rightarrow2, K',\downarrow}$& 0.061928\\
         $\Delta_{-1\rightarrow2, K,\downarrow}$
& $\Delta_{-1\rightarrow2, K',\uparrow}$
& 0.063917\\
        \bottomrule
    \end{tabular}
    \label{vspmtrans}
\end{table}

\begin{table}[h]
    \centering
    \caption{Table of M-O excitation energies for allowed transitions in the $n=(-2,-1,0,1,2)$ subspace for $B=0.5$ T in BI as shown in Fig. \ref{sigmasilicene}(e,f) and Fig. \ref{transsili}(e,f).}
    \label{trans5}
    \begin{tabular}{ccc}
        \toprule
         $\Delta_{m\rightarrow n, K,\uparrow(\downarrow)}$& $\Delta_{m\rightarrow n, K',\uparrow(\downarrow)}$& Photonic energy (eV)\\
        \midrule
         $\Delta_{0\rightarrow1, K,\uparrow}$ 
& $\Delta_{-1\rightarrow0, K',\downarrow}$& 0.0298463\\
         $\Delta_{0\rightarrow1, K,\downarrow}$& $\Delta_{-1\rightarrow0, K',\uparrow}$& 0.0398938\\
         $\Delta_{-1\rightarrow2, K,\uparrow}$& $\Delta_{-1\rightarrow2, K',\downarrow}$& 0.062432\\
         $\Delta_{-1\rightarrow2, K,\downarrow}$
& $\Delta_{-1\rightarrow2, K',\uparrow}$
& 0.0663106\\
        \bottomrule
    \end{tabular}
    \label{bitrans}
\end{table}

 \subsubsection{Magneto-optical Response of silicene}

Now, we explore the M-O conductivities for different values of finite staggered sub-lattice potential and SOC that results in the different TI phases of silicene\cite{tabert2013magneto,PhysRevB.101.205408,tahir2015magneto,Khan:25,shah2019magneto}. A static electric field tunes the electronic band structure of the staggered graphene family by inducing $M$ which is then responsible for spin- and valley-polarized responses, playing a crucial role in controlling M-O excitations. Fig. \ref{sigmasilicene} shows conductivity spectra, where we can observe multiple absorption peaks, the lower energy ones, originating from LLLs and higher energy excitations occurring due to higher LLs. It is evident that the transitions involving LLL exhibit stronger intensity compared to higher LL transitions, dominating the conductivity spectrum and defining the topology of the phase as well. Fig. \ref{transsili} shows the corresponding excitation energies for the allowed transitions in the subspace $n= (-2,-1,0,1,2)$ and Tables (\ref{qshitrrans}, \ref{vspmtrans} and \ref{bitrans}) give values of excitation energies $\Delta_{m\rightarrow n,K(K'),\uparrow(\downarrow)}$ for different TI regimes triggered by variation of $M$ compared to $\Delta_{so}$.

Initially, when $M<\Delta_{so}$, the system is in QSHI regime. Fig. \ref{sigmasilicene}(a,b) show the absorptive M-O conductivity peaks in $K$ and $K'$ valleys, respectively, for $M=0.5\Delta_{so}$ and $\mu_F=0$. The first peak occurs due to excitation at an energy $\Delta_{-1\rightarrow0,K,\uparrow}$, while, the spin-down peak is blue shifted to $\Delta_{0\rightarrow1,K,\downarrow}$. In $K'$ valley, the first peak occurs at $\Delta_{0\rightarrow1,K',\downarrow}$ and the second peak due to spin-up is blue-shifted to $\Delta_{-1\rightarrow0,K',\uparrow}$. The higher LL excitations are much shorter and show quite similar behavior in both valleys. The M-O excitation frequencies in QSHI regime are listed in Table \ref{qshitrrans} and the corresponding transitions are shown explicitly in Fig. \ref{transsili}(a,b).

An increase in the electric field makes inter-band spin-polarized peaks to split further apart and finally triggers transition to metallic phase called VSPM state at $M=\Delta_{so}$\cite{PhysRevLett.107.076802} where the gap-closing occurs for one of the spin-split bands. Fig. \ref{sigmasilicene}(c,d) plots the longitudinal conductivity versus photon energy for VSPM phase. The lowest frequency peaks move apart: the $\Delta_{-1\rightarrow 0,K,\uparrow}$ peak is red-shifted and the $\Delta_{0\rightarrow 1,K,\downarrow}$ peak is blue-shifted. This shifting of spin-down absorption peaks to lower excitation energies (red-shift) indicates closing of the smallest band gap for Dirac fermions. The allowed transitions in VSPM regime are given in Table \ref{vspmtrans}  and the corresponding transitions are shown explicitly in Fig. \ref{transsili}(c,d).

Further increasing $M$ leads to re-opening of the gaps and transition of our system from the VSPM to (BI) state. In the BI regime, all inter-band absorption peaks move to higher energies due to the widening of the band gap as illustrated in Fig.~\ref{sigmasilicene}(e,f). The magneto-excitation frequencies in Table \ref{bitrans} for the first two transitions show that $\Delta_{0\rightarrow 1,K,\uparrow(\downarrow)}(\Delta_{-1\rightarrow 0,K',\downarrow(\uparrow)})$ are restricted to $K(K')$ valley. Fig. \ref{transsili}(e,f) show the M-O excitations in quantized LLs. The reappearance of the optical transition $\Delta_{0\rightarrow 1,K,\uparrow(\downarrow)}$ in the valley $K$ provides clear evidence of the transition from TI phase to trivial phase.

TI phases and externally driven phase transitions in silicene, induced by perpendicular electric fields \cite{Bampoulis202023quantum,Khan:25,PhysRevB.101.205408}, off-resonant circularly polarized light \cite{duan2018bulk,Khan:25}, and exchange fields \cite{ezawa2013spin} competing with intrinsic SOC \cite{shah2019magneto} have been extensively explored over the past two decades. However, a unified topological description of M-O responses across these symmetry-breaking mechanisms has remained incomplete. In a seminal work, Tabert and Nicol (2013) systematically investigated the M-O response of silicene using the Kubo formalism \cite{PhysRevLett.110.197402,tabert2013magneto}.  Ref. \cite{tabert2013magneto} presented a detailed study of M-O response of Silicene as function of chemical potential $\mu_F$ where $\mu_F$ is fixed first between $n=0$ and $n=1$ and subsequently between $n=1$ and $n=2$. This setup differs fundamentally from the present work where $\mu_F$ is fixed at $E=0$ with the specific objective of identifying universal M-O signature that distinguish TI phase from trivial BI phase.

Our results establish the modified Haldane model as a universal framework for describing the M–O response of Dirac materials. While excitation energies of each Kramers pair are identical, fixing $\mu=0$ allows only one of the $0\rightarrow1$ or $-1\rightarrow0$ transitions due to LLL inversion controlled by $M$ and $\Delta_{so}$. For the trivial Kramers pair $(K,\downarrow)$ and $(K',\uparrow)$, the transitions $\Delta_{0\rightarrow1,K,\downarrow}$ and $\Delta_{-1\rightarrow0,K',\uparrow}$ persist throughout the parameter space. In contrast, the topologically important Kramers pair exhibits phase-dependent selection rules: $\Delta_{0\rightarrow1,K,\uparrow}$ and $\Delta_{-1\rightarrow0,K',\downarrow}$ in the trivial phase, and $\Delta_{0\rightarrow1,K',\downarrow}$ and $\Delta_{-1\rightarrow0,K,\uparrow}$ in the TI phase. This switching constitutes a universal optical fingerprint of the topological phase transition.

\subsection{Transition metal dichalcogenides(TMDCs)}
Mono-layer TMDCs (i.e. $MX_2, M = Mo, W; X = S, Se$)  \cite{geim2013van,nature2014,mak2016photonics} is characterized by a wide band gap extending from the near-infrared to the visible spectral range. Their large intrinsic band gap (hundreds of meV) keeps them in a conventional band-insulating phase, but they still exhibit topological effects in a broader, valley-resolved sense\cite{PhysRevB.109.165441}. Valley-dependent inter-LL optical transitions have been reported in \cite{chu2014valley}, in contrast to graphene where such transitions are valley independent, although both systems exhibit valley-contrasting Berry physics.  The presence of inequivalent valleys hosting opposite spin polarizations, such that the state $(K,\uparrow)$ is coupled to $(K',\downarrow)$ 
and vise versa makes them significantly interesting \cite{PhysRevB.101.045424}.

Magneto-transport in monolayer TMDCs has been widely studied since last decade \cite{chu2014valley,PhysRevB.101.045424,PhysRevB.109.165441,Xiao2012,PhysRevLett.113.077201,PhysRevB.93.035406,PhysRevB.94.045415,PhysRevB.92.125303}. Nguyen et al explored the LLs spectrum, M-O conductivities of different TMDCs taking into account material dependence of parameters and the band-dependence of the SOC strength and also including the spin and valley Zeeman couplings \cite{PhysRevB.101.045424}. However, these couplings enter as additive energy shifts and primarily induce spin and valley splittings in the spectrum but do not alter the band topology. In contrast, application of strain leads to the emergence of QSHI by producing a pseudo magnetic field which has opposite sign in different valleys to preserve TRS as shown in \cite{qian2014quantum,PhysRevLett.113.077201}. The structural distortion causes intrinsic band inversion and modifies Dirac mass term driving topological phase transition. The topological character in TMDCs is governed by valley- and spin-dependent Dirac mass terms hence, the other terms affect only the LL spectrum and optical response.

Here, we retain only the minimal terms, that are sufficient to capture the essential topological features of TMDCs, while neglecting additional contributions. We have taken $ \phi = 5\pi/6(-\pi/6)$ for spin-up (spin-down) due to the asymmetry between VB and CB but we have ignored the band dependence of SOC. Adjusting $M = \Delta/2 , t_2 = \Delta_{tmdc}/3 \sqrt 3,\text{ and } \phi = 5\pi/6(-\pi/6),$ one can recover the dispersion relation, characteristic of monolayer materials $MX_2$ in the valleys $K$ and $K'$ for both spins as
\begin{widetext}
\begin{align}
    E_{n,\eta}^{\tau,s} &= 
\begin{cases}
s\Delta_{tmdc}/2+\eta \sqrt{ 2v_F^2\hbar eB|n| + 1/4(\Delta-s\tau\Delta_{tmdc})^2}, & n \neq 0,\\
-\tau \Delta/2+s\Delta_{tmdc}, & n=0.
\end{cases} 
\end{align}
\end{widetext}

 The structure of the LLs in TMDC-like systems reflects the valley-contrasting spin splitting originating from broken inversion symmetry and strong SOC. As a result, the LLs exhibit complete spin polarization, but this polarization appears in different bands for the two inequivalent valleys. The LLL is strongly spin-polarized and appears only in the CB at the $K'$ valley and in the VB at the $K$ valley, reflecting the valley-selective spin splitting. The sign of splitting is opposite at the two inequivalent valleys, with  magnitude equal to $2\Delta_{tmdc}$. Thus, the LLL exists only in the band where the intrinsic valley-dependent spin splitting occurs.  In each case, the LLL follows the valley-dependent intrinsic spin ordering, so that the LLL is restricted to the band in which the spin splitting is present. The LLL states have energies: 

\begin{figure}[ht!]
        \centering
    \begin{tabular}{c}
      \includegraphics[width=0.8\columnwidth]{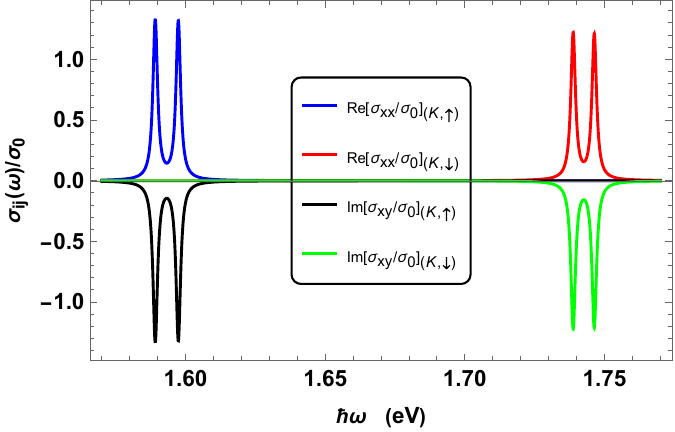} \\
    {\small\textbf{(a)}} \\[0.8em]

    \includegraphics[width=0.8\columnwidth]{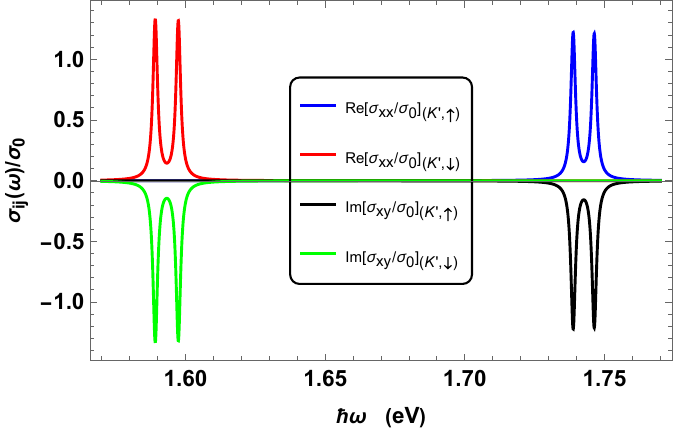} \\
    {\small\textbf{(b)}}
    \end{tabular}
\caption{(a,b) M-O conductivities i.e. $\mathrm{Re[\sigma_{xx}/\sigma_0]}$ and $\mathrm{Im[\sigma_{xy}/\sigma_0]}$ at $K$ and $K'$ valley respectively. The blue (red) curves represent $\mathrm{Re[\sigma_{xx}/\sigma_0]}$ for up (down) spins and black (green) curves represent $\mathrm{Im[\sigma_{xy}/\sigma_0]}$ for up (down) spins. In $K$ valley, the first peak corresponds to $\Delta_{0\rightarrow1}$ for spin-up. For spin-down in the same valley, the peaks are blue shifted to energy. For $K'$ valley, $\Delta_{0\rightarrow1}$ is replaced with $\Delta_{-1\rightarrow0}$ because of shift of LL from VB to CB. We have taken $B=5$ T and $\Gamma=1$ meV.}
    \label{tmdcsigma}
\end{figure}

The band gap is large $E_0^{K',\downarrow}-E_0^{K,\uparrow}=\Delta-2\Delta_{tmdc}$  where $\Delta$ is of the order of eV. While the spin splitting in both valleys is equal to $2 \Delta_{tmdc}$ which is of the order of meV, so, to clearly resolve the higher LLs separately, the CB and the VB  have been explicitly plotted in Fig. \ref{tmdc}(a,b). For higher LLs; $(n>1)$, the behavior is more subtle. In the spin-split valley, i.e., $K'$ valley in CB and $K$ valley in VB, the higher LLs remain spin-polarized, following the same valley-selective spin ordering as LLL. However, in the opposite valley ($K$ in CB and $K'$ in VB), the higher LLs in that band become nearly spin-degenerate because the spin splitting is extremely small for $n>1$. These nearly spin-degenerate LLs lie inside the energy gap produced by the spin-polarized LLLs of the opposite valley. More precisely, the approximate spin splitting is $E_{n\geq1}^{\tau,\uparrow}-E_{n\geq1}^{\tau,\downarrow}\approx\frac{ 4v_F^2\hbar eB}{(\Delta^2-\Delta_{tmdc}^2)} \Delta_{tmdc}\ll \Delta_{tmdc}$; this confirms the approximate spin degeneracy of higher LLs.

Consequently, spin-polarized LL ladders appear only in the band where the valley's intrinsic spin splitting is active, and spin-degenerate LL ladders appear in the same band but in the opposite valley, positioned within the gap opened by the spin-polarized LLLs. This creates a valley-asymmetric LL spectrum, where one valley hosts a sequence of spin-polarized LLLs in a given band while the opposite valley hosts nearly spin-degenerate LLs in the same energy window, positioned within the spin-splitting gap opened by the polarized valley. This interplay yields a characteristic LL pattern: the LLL is fully valley- and spin-polarized, while the higher LLs exhibit mixed behavior, being polarized in one valley and nearly degenerate in the other, with the degenerate ones occurring inside the spin-splitting gap of the polarized valley \cite{PhysRevB.101.045424}.

Fig. \ref{tmdc}(c,d) represent the DOS $(D(E)/D_0)$ for the CB and VB up to $n=2$. The leftmost green and black peaks in (c) and the rightmost blue and red peaks in (d) correspond to the LLL. The intermediate peaks arise from spin degenerate LLs associated with unpolarized valleys.  Although the LLL resembles that of a trivial BI, the higher levels exhibit pronounced spin-valley polarization leading to valley dependent M-O transitions. This behavior places TMDCs within the same unified theoretical framework as Xenes and graphene-based systems.

\begin{table}[h]
    \centering
    \caption{Table of M-O excitation energies for the allowed transitions in the $n=(-2,-1,0,1,2)$ subspace for $B=5$ T in TMDC as shown in Fig. \ref{tmdcsigma}(a,b).}
    \label{trans2}
    \begin{tabular}{ccc}
        \toprule
         $\Delta_{m\rightarrow n, K,\uparrow(\downarrow)}$& $\Delta_{m\rightarrow n, K',\uparrow(\downarrow)}$& Photonic energy (eV)\\
        \midrule
         $\Delta_{0\rightarrow1, K,\uparrow}$ 
& $\Delta_{-1\rightarrow0, K',\downarrow}$& 1.58914 \\
         $\Delta_{-1\rightarrow2, K,\uparrow}$  
& $\Delta_{-1\rightarrow2, K',\downarrow}$  
& 1.5974 \\
         $\Delta_{0\rightarrow1, K,\downarrow}$ 
& $\Delta_{-1\rightarrow0, K',\uparrow}$ 
& 1.73878\\
         $\Delta_{-1\rightarrow2, K,\downarrow}$
& $\Delta_{-1\rightarrow2, K',\uparrow}$
& 1.74634 \\
        \bottomrule
    \end{tabular}
    \label{tmdctrans}
\end{table}

Fig. \ref{tmdcsigma}(a,b) present the real and imaginary parts of $\mathrm{\sigma_{xx}}$ and $\mathrm{\sigma_{xy}}$ of monolayer $MoS_2$, for spin-up and spin-down electrons in the $K$ and $K'$ valleys, respectively. Multiple absorption peaks appear at well-defined magneto-excitation energies arising from electron transitions between different LLs for both spin-up and spin-down channels. Two distinct jumps are observed in the conductivity spectra plotted in Fig. \ref{tmdcsigma}(a) and (b) and the excitation energies for allowed transitions are listed in Table \ref{tmdctrans} \cite{PhysRevB.89.115413}. 
\\
In $K(K')$ valley, the first peak corresponds to \(\Delta_{0\rightarrow 1} \)( \(\Delta_{-1\rightarrow 0} \)) while the second peak corresponds to  \(\Delta_{-1\rightarrow2} \). The higher LL transitions are peaked at higher photonic energies. The blue (red) lines represent $\mathrm{Re[\sigma_{xx}/\sigma_0]}$ for spin-up (down) and black (green) lines represent $\mathrm{Im[\sigma_{xy}/\sigma_0]}$ for spin-up (down) \cite{Schwingenschlögl}. Explicit values of the first magneto-excitation energies corresponding to the $\Delta_{0\rightarrow 1,K,\uparrow(\downarrow)}$ and $\Delta_{-1\rightarrow 0,K',\downarrow(\uparrow)}$ for spin-up and spin-down are summarized in Table \ref{tmdctrans}. spin-down (up) electrons experience a much larger band gap in contrast to spin-up (down) electrons in $K(K')$ valley leading to excitations at higher photon energies. 

\section{Conclusion}\label{con}

In conclusion, we have theoretically examined quantum magnetotransport and M-O responses in a broad class of monolayer hexagonal 2D systems, including Xenes and TMDCs, under a perpendicular magnetic field using a unified and tunable generic Hamiltonian. By varying material-specific parameters, this framework captures distinct TI phases and their transitions, enabling a direct comparison between trivial BI, topological Dirac systems, and valley-polarized semiconductors, while separating universal M-O features from material-dependent details.

Employing the Kubo formalism, we show that inter-LL optical transitions generate pronounced resonance peaks in the longitudinal and Hall conductivities, which serve as reliable fingerprints of different TI regimes and their field-driven transitions. The valley-resolved M-O response further provides a clear distinction between trivial BI and TI phases.

In this context, the M-O transition $\Delta_{0\rightarrow1}$ serves as a direct spectroscopic indicator of the underlying TI phase. When the chemical potential is fixed at charge neutrality, the survival of the $\Delta_{0\rightarrow1}$  transition depends on the location of the LLL, which is controlled by the sign of the Dirac mass. In TI (BI) regime, the LLL becomes spin-band (valley-band) polarized. As a result, the allowed transition $\Delta_{0\rightarrow1}$ occurs for the spin-down sector (K valley). The corresponding transition for the opposite spin or valley is Pauli blocked. Thus, the sector in which the  transition appears: spin-resolved in the TI phase versus valley-resolved in the BI phase, provides a clear M-O fingerprint distinguishing the TI phase from the trivial BI one.

This model effectively captures distinct TI phases of silicene family and transitions between them. TMDCs, on the other hand, lack band inversion and exhibit a LLL spectrum characteristic of a trivial BI, their higher Landau levels display strong spin-valley locking. This places TMDCs naturally within the unified Modified-Haldane-model description, where their M-O response represents a distinct manifestation of the same underlying topology.

\renewcommand{\bibname}{References}

\section*{References}

\bibliography{refr}

@article{xiao2010berry,
title={Berry phase effects on electronic properties},
author={Xiao, Di and Chang, Ming-Che and Niu, Qian},
journal={Reviews of modern physics},
volume={82},
number={3},
pages={1959--2007},
year={2010},
publisher={APS},
doi={10.1103/RevModPhys.82.1959}
}

@article{Bampoulis202023quantum,
  title = {Quantum Spin Hall States and Topological Phase Transition in Germanene},
  author = {Bampoulis, Pantelis and Castenmiller, Carolien and Klaassen, Dennis J. and van Mil, Jelle and Liu, Yichen and Liu, Cheng-Cheng and Yao, Yugui and Ezawa, Motohiko and Rudenko, Alexander N. and Zandvliet, Harold J. W.},
  journal = {Phys. Rev. Lett.},
  volume = {130},
  issue = {19},
  pages = {196401},
  numpages = {6},
  year = {2023},
  month = {May},
  publisher = {American Physical Society},
  doi = {10.1103/PhysRevLett.130.196401},
  url = {https://link.aps.org/doi/10.1103/PhysRevLett.130.196401}
}

@article{pesin2012spin,
title={Spintronics in monolayer graphene},
author={Pesin, D. and MacDonald, A. H.},
journal={Nature Materials},
volume={11},
pages={409--416},
year={2012},
publisher={Nature Publishing Group},
doi={10.1038/nmat3305}
}

@article{ezawa2013spin,
title={Spin valleytronics in silicene: Quantum spin Hall--quantum anomalous Hall insulators and single-valley semimetals},
author={Ezawa, Motohiko},
journal={Physical Review B},
volume={87},
number={15},
pages={155415},
year={2013},
publisher={APS},
doi={10.1103/PhysRevB.87.155415}
}

@article{tabert2013magneto,
title={Magneto-optical conductivity of silicene and other buckled honeycomb lattices},
author={Tabert, Calvin J and Nicol, Elisabeth J},
journal={Physical Review B},
volume={88},
number={8},
pages={085434},
year={2013},
publisher={APS},
doi={10.1103/PhysRevB.88.085434}
}

@article{CHOI2017116,
title = {Recent development of two-dimensional transition metal dichalcogenides and their applications},
journal = {Materials Today},
volume = {20},
number = {3},
pages = {116-130},
year = {2017},
issn = {1369-7021},
doi = {https://doi.org/10.1016/j.mattod.2016.10.002},
url = {https://www.sciencedirect.com/science/article/pii/S1369702116302917},
author = {Wonbong Choi and Nitin Choudhary and Gang Hee Han and Juhong Park and Deji Akinwande and Young Hee Lee}
}

@article{ezawa2012topological,
title={Topological phase transition and electrically tunable diamagnetism in silicene},
author={Ezawa, Motohiko},
journal={The European Physical Journal B},
volume={85},
pages={1--5},
year={2012},
doi={10.1140/epjb/e2012-30577-0},
publisher={Springer}
}

@article{PhysRevB.107.235417,
  title = {Magnetoresistance in noncentrosymmetric two-dimensional systems},
  author = {Faridi, Azadeh and Asgari, Reza},
  journal = {Phys. Rev. B},
  volume = {107},
  issue = {23},
  pages = {235417},
  numpages = {9},
  year = {2023},
  month = {Jun},
  publisher = {American Physical Society},
  doi = {10.1103/PhysRevB.107.235417},
  url = {https://link.aps.org/doi/10.1103/PhysRevB.107.235417}
}

@article{qian2014quantum,
  title={Quantum spin Hall effect in two-dimensional transition metal dichalcogenides},
  author={Qian, Xiaofeng and Liu, Junwei and Fu, Liang and Li, Ju},
  journal={Science},
  volume={346},
  number={6215},
  pages={1344--1347},
  year={2014},
doi = {10.1126/science.1256815},
  publisher={American Association for the Advancement of Science}
}

@article{shah2019magneto,
title={Magneto-optical effects in the Landau level manifold of 2D lattices with spin-orbit interaction},
author={Shah, Muhammad and Anwar, Muhammad Sajid},
journal={Optics Express},
volume={27},
number={16},
pages={23217--23233},
year={2019},
publisher={OSA},
doi={10.1364/OE.27.023217}
}

@article{tahir2015magneto,
title={Magneto-optical transport properties of monolayer phosphorene},
author={Tahir, M and Vasilopoulos, P and Peeters, F M},
journal={Physical Review B},
volume={92},
pages={045420},
year={2015},
doi={10.1103/PhysRevB.92.045420}
}

@article{gusynin2005unconventional,
title={Unconventional integer quantum Hall effect in graphene},
author={Gusynin, V P and Sharapov, S G},
journal={Physical Review Letters},
volume={95},
number={14},
pages={146801},
year={2005},
publisher={American Physical Society},
doi={10.1103/PhysRevLett.95.146801}
}

@article{chu2014valley,
title={Valley-splitting and valley-dependent inter-Landau-level optical transitions in monolayer MoS2 quantum Hall systems},
author={Chu, Rong-Ling and Li, Xiao and Wu, Shen and Niu, Qian and Yao, Wang and Xu, Xiaodong and Zhang, Chuanwei},
journal={Physical Review B},
volume={90},
pages={045427},
year={2014},
doi={10.1103/PhysRevB.90.045427}
}

@article{yuan2018chiral,
title={Chiral Landau levels in Weyl semimetal NbAs with multiple topological carriers},
author={Yuan, Xiang and Yan, Cheng and Song, Chaoyu and Zhang, Mengyao and Li, Zhong and Zhang, Cheng and Liu, Yuan and Wang, Weiyi and Zhao, Mingliang and Lin, Zhilin and Xie, Tao and Ludwig, J and Jiang, Yadong and Zhang, Xinxin and Shang, Chun and Ye, Zhizhen and Wang, Jian and Chen, Faxian and Xia, Zhengcai and Smirnov, Dmitry},
journal={Nature Communications},
volume={9},
pages={1854},
year={2018},
doi={10.1038/s41467-018-04163-4}
}

@article{li2013magneto,
title={Magneto-optical conductivity in a topological insulator},
author={Li, Zhi and Carbotte, J P},
journal={Physical Review B},
volume={88},
pages={045414},
year={2013},
doi={10.1103/PhysRevB.88.045414}
}

@article{stern1967properties,
title={Properties of semiconductor surface inversion layers in the electric quantum limit},
author={Stern, Frank and Howard, W E},
journal={Physical Review},
volume={163},
pages={816--835},
year={1967},
doi={10.1103/PhysRev.163.816}
}

@article{novoselov2005two,
title={Two-dimensional gas of massless Dirac fermions in graphene},
author={Novoselov, K S and Geim, A K and Morozov, S V and Jiang, D and Katsnelson, M I and Grigorieva, I V and Dubonos, S V and Firsov, A A},
journal={Nature},
volume={438},
pages={197--200},
year={2005},
doi={10.1038/nature04233}
}

@article{zhang2005experimental,
title={Experimental observation of the quantum Hall effect and Berry's phase in graphene},
author={Zhang, Yuanbo and Tan, Yan-Wen and Stormer, Horst L and Kim, Philip},
journal={Nature},
volume={438},
pages={201--204},
year={2005},
doi={10.1038/nature04235}
}

@article{klitzing1986quantized,
title={The quantized Hall effect},
author={Klitzing, Klaus von},
journal={Reviews of Modern Physics},
volume={58},
pages={519--531},
year={1986},
doi={10.1103/RevModPhys.58.519}
}

@article{PhysRevLett.125.046403,
  title = {Unraveling the Topological Phase of ${\mathrm{ZrTe}}_{5}$ via Magnetoinfrared Spectroscopy},
  author = {Jiang, Y. and Wang, J. and Zhao, T. and Dun, Z. L. and Huang, Q. and Wu, X. S. and Mourigal, M. and Zhou, H. D. and Pan, W. and Ozerov, M. and Smirnov, D. and Jiang, Z.},
  journal = {Phys. Rev. Lett.},
  volume = {125},
  issue = {4},
  pages = {046403},
  numpages = {6},
  year = {2020},
  month = {Jul},
  publisher = {American Physical Society},
  doi = {10.1103/PhysRevLett.125.046403},
  url = {https://link.aps.org/doi/10.1103/PhysRevLett.125.046403}
}

@article{PhysRevB.102.235134,
  title = {Magneto-optical conductivity in generic Weyl semimetals},
  author = {St\aa{}lhammar, Marcus and Larana-Aragon, Jorge and Knolle, Johannes and Bergholtz, Emil J.},
  journal = {Phys. Rev. B},
  volume = {102},
  issue = {23},
  pages = {235134},
  numpages = {14},
  year = {2020},
  month = {Dec},
  publisher = {American Physical Society},
  doi = {10.1103/PhysRevB.102.235134},
  url = {https://link.aps.org/doi/10.1103/PhysRevB.102.235134}
}

@article{PhysRevB.105.195102,
  title = {Nonlinear optical response of type-II Weyl fermions in two dimensions},
  author = {Tamashevich, Yaraslau and Villari, Leone Di Mauro and Ornigotti, Marco},
  journal = {Phys. Rev. B},
  volume = {105},
  issue = {19},
  pages = {195102},
  numpages = {7},
  year = {2022},
  month = {May},
  publisher = {American Physical Society},
  doi = {10.1103/PhysRevB.105.195102},
  url = {https://link.aps.org/doi/10.1103/PhysRevB.105.195102}
}

@article{ando1982electronic,
title={Electronic properties of two-dimensional systems},
author={Ando, Tsuneya and Fowler, Alan B and Stern, Frank},
journal={Reviews of Modern Physics},
volume={54},
number={2},
pages={437--672},
year={1982},
publisher={American Physical Society},
doi={10.1103/RevModPhys.54.437}
}

@book{giuliani2005quantum,
title={Quantum Theory of the Electron Liquid},
author={Giuliani, Gabriele F and Vignale, Giovanni},
publisher={Cambridge University Press},
year={2005},
doi={10.1017/CBO9780511619915}
}

@article{castro2009electronic,
title={The electronic properties of graphene},
author={Castro Neto, A H and Guinea, F and Peres, N M R and Novoselov, K S and Geim, A K},
journal={Reviews of Modern Physics},
volume={81},
number={1},
pages={109--162},
year={2009},
publisher={American Physical Society},
doi={10.1103/RevModPhys.81.109}
}

@article{goerbig2011electronic,
title={Electronic properties of graphene in a strong magnetic field},
author={Goerbig, M O},
journal={Reviews of Modern Physics},
volume={83},
number={4},
pages={1193--1243},
year={2011},
publisher={American Physical Society},
doi={10.1103/RevModPhys.83.1193}
}

@article{Novoselov2004,
  title={Electric field effect in atomically thin carbon films},
  author={Novoselov, Kostya S and Geim, Andre K and Morozov, Sergei V and Jiang, De-eng and Zhang, Yanshui and Dubonos, Sergey V and Grigorieva, Irina V and Firsov, Alexandr A},
  journal={Science},
  volume={306},
  number={5696},
  pages={666--669},
  year={2004},
  doi={10.1126/science.11028},
  publisher={American Association for the Advancement of Science}
}

@book{vanderbilt2018berry,
  title={Berry phases in electronic structure theory: electric polarization, orbital magnetization and topological insulators},
  author={Vanderbilt, David},
  year={2018},
  publisher={Cambridge University Press}
}

@article{cooper2019topological,
  title={Topological bands for ultracold atoms},
  author={Cooper, NR and Dalibard, J and Spielman, IB},
  journal={Reviews of Modern Physics},
  volume={91},
  number={1},
  pages={015005},
  year={2019},
  doi={10.1103/RevModPhys.91.015005},
  publisher={APS}
}

@article{ren2016topological,
  title={Topological phases in two-dimensional materials: a review},
  author={Ren, Yafei and Qiao, Zhenhua and Niu, Qian},
  journal={Reports on Progress in Physics},
  volume={79},
  number={6},
  pages={066501},
  year={2016},
  doi={10.1088/0034-4885/79/6/066501},
  publisher={IOP Publishing}
}

@article{pratama2020circular,
  title={Circular dichroism and Faraday and Kerr rotation in two-dimensional materials with intrinsic {Hall} conductivities},
  author={Pratama, FR and Ukhtary, M Shoufie and Saito, Riichiro},
  journal={Physical Review B},
  volume={101},
  number={4},
  pages={045426},
  year={2020},
  doi={10.1103/PhysRevB.101.045426},
  publisher={APS}
}

@article{ezawa2012spin,
  title={Spin-valley optical selection rule and strong circular dichroism in silicene},
  author={Ezawa, Motohiko},
  journal={Physical Review B},
  volume={86},
  number={16},
  pages={161407},
  year={2012},
  doi={10.1103/PhysRevB.86.161407},
  publisher={APS}
}

@article{xu2013large,
  title={Large-gap quantum spin {Hall} insulators in tin films},
  author={Xu, Yong and Yan, Binghai and Zhang, Hai-Jun and Wang, Jing and Xu, Gang and Tang, Peizhe and Duan, Wenhui and Zhang, Shou-Cheng},
  journal={Physical Review Letters},
  volume={111},
  number={13},
  pages={136804},
  year={2013},
  doi={10.1103/PhysRevLett.111.136804},
  publisher={APS}
}

@article{geim2013van,
  title={Van der Waals heterostructures},
  author={Geim, Andre K and Grigorieva, Irina V},
  journal={Nature},
  volume={499},
  number={7459},
  pages={419--425},
  year={2013},
  doi={10.1038/nature12385},
  publisher={Nature Publishing Group UK London}
}

@article{mak2016photonics,
  title={Photonics and optoelectronics of 2D semiconductor transition metal dichalcogenides},
  author={Mak, Kin Fai and Shan, Jie},
  journal={Nature Photonics},
  volume={10},
  number={4},
  pages={216--226},
  year={2016},
  doi={10.1038/nphoton.2015.282},
  publisher={Nature Publishing Group UK London}
}

@article{PhysRevB.89.115413,
  title = {Intrinsic optical conductivity of modified Dirac fermion systems},
  author = {Rostami, Habib and Asgari, Reza},
  journal = {Phys. Rev. B},
  volume = {89},
  issue = {11},
  pages = {115413},
  numpages = {12},
  year = {2014},
  month = {Mar},
  publisher = {American Physical Society},
  doi = {10.1103/PhysRevB.89.115413},
  url = {https://link.aps.org/doi/10.1103/PhysRevB.89.115413}
}

@article{schaibley2016valleytronics,
  title={Valleytronics in 2D materials},
  author={Schaibley, John R and Yu, Hongyi and Clark, Genevieve and Rivera, Pasqual and Ross, Jason S and Seyler, Kyle L and Yao, Wang and Xu, Xiaodong},
  journal={Nature Reviews Materials},
  volume={1},
  number={11},
  pages={1--15},
  year={2016},
  doi={Valleytronics in 2D materials},
  publisher={Nature Publishing Group}
}

@article{pesin2012spintronics,
  title={Spintronics and pseudospintronics in graphene and topological insulators},
  author={Pesin, Dmytro and MacDonald, Allan H},
  journal={Nature Materials},
  volume={11},
  number={5},
  pages={409--416},
  year={2012},
  doi={10.1038/nmat3305},
  publisher={Nature Publishing Group UK London}
}

@article{sasaki2008pseudospin,
  title={Pseudospin and deformation-induced gauge field in graphene},
  author={Sasaki, Ken-ichi and Saito, Riichiro},
  journal={Progress of Theoretical Physics Supplement},
  volume={176},
  pages={253--278},
  year={2008},
  doi={10.1143/PTPS.176.253},
  publisher={Oxford University Press}
}

@article{Wang_2014,
doi = {10.1088/1367-2630/16/4/045015},
url = {https://doi.org/10.1088/1367-2630/16/4/045015},
year = {2014},
month = {apr},
publisher = {IOP Publishing},
volume = {16},
number = {4},
pages = {045015},
author = {Wang, S K and Wang, J and Chan, K S},
title = {Multiple topological interface states in silicene},
journal = {New Journal of Physics}
}

@article{duan2018bulk,
  title={Bulk RKKY signatures of topological phase transition in silicene},
  author={Duan, Hou-Jian and Wang, Chen and Zheng, Shi-Han and Wang, Rui-Qiang and Pan, Da-Ru and Yang, Mou},
  journal={Scientific reports},
  volume={8},
  number={1},
  pages={6185},
  year={2018},
  publisher={Nature Publishing Group UK London},
  url = {https://doi.org/10.1038/s41598-018-24567-w},
doi = {10.1038/s41598-018-24567-w},
}

@article{Khan:25,
author = {Imtiaz Khan and Mudasir Shah and Muzamil Shah},
journal = {Opt. Mater. Express},
number = {6},
pages = {1294--1306},
publisher = {Optica Publishing Group},
title = {Quantum transport in buckled Xene materials},
volume = {15},
month = {Jun},
year = {2025},
url = {https://opg.optica.org/ome/abstract.cfm?URI=ome-15-6-1294},
doi = {10.1364/OME.568069},
}

@article{PhysRevLett.107.076802,
  title = {Quantum Spin Hall Effect in Silicene and Two-Dimensional Germanium},
  author = {Liu, Cheng-Cheng and Feng, Wanxiang and Yao, Yugui},
  journal = {Phys. Rev. Lett.},
  volume = {107},
  issue = {7},
  pages = {076802},
  numpages = {4},
  year = {2011},
  month = {Aug},
  publisher = {American Physical Society},
  doi = {10.1103/PhysRevLett.107.076802},
  url = {https://link.aps.org/doi/10.1103/PhysRevLett.107.076802}
}

@article{PhysRevLett.98.157402,
  title = {Anomalous Absorption Line in the Magneto-Optical Response of Graphene},
  author = {Gusynin, V. P. and Sharapov, S. G. and Carbotte, J. P.},
  journal = {Phys. Rev. Lett.},
  volume = {98},
  issue = {15},
  pages = {157402},
  numpages = {4},
  year = {2007},
  month = {Apr},
  publisher = {American Physical Society},
  doi = {10.1103/PhysRevLett.98.157402},
  url = {https://link.aps.org/doi/10.1103/PhysRevLett.98.157402}
}

@article{Haldane1988,
  title = {Model for a Quantum Hall Effect without Landau Levels: Condensed-Matter Realization of the "Parity Anomaly"},
  author = {Haldane, F. D. M.},
  journal = {Phys. Rev. Lett.},
  volume = {61},
  issue = {18},
  pages = {2015--2018},
  numpages = {0},
  year = {1988},
  month = {Oct},
  publisher = {American Physical Society},
  doi = {10.1103/PhysRevLett.61.2015},
  url = {https://link.aps.org/doi/10.1103/PhysRevLett.61.2015}
}

@article{Schwingenschlögl,
  title = {Valley polarized quantum Hall effect and topological insulator phase transitions in silicene},
  author = {Schwingenschlögl, U. and Tahir, M},
  journal = {Scientific Reports},
  volume = {3},
  issue = {1},
  year = {2013},
  publisher = {nature},
  doi = { 10.1038/srep01075},
  url = {https://www.nature.com/articles/srep01075}
}

@article{buckled,
  title = {Realising Haldane's vision for a Chern insulator in buckled lattices},
  author = {Wright, Anthony, R},
  journal = {Scientific Reports},
  volume = {3},
  issue = {1},
  year = {2013},
  publisher = {nature},
  doi = {10.1038/srep02736},
  url = {https://doi.org/10.1038/srep02736}
}

@article{doi:10.7566/JPSJ.84.121003,
author = {Ezawa ,Motohiko},
title = {Monolayer Topological Insulators: Silicene, Germanene, and Stanene},
journal = {Journal of the Physical Society of Japan},
volume = {84},
number = {12},
pages = {121003},
year = {2015},
doi = {10.7566/JPSJ.84.121003},
URL = {https://doi.org/10.7566/JPSJ.84.121003},
eprint = {https://doi.org/10.7566/JPSJ.84.121003}
}

@article{PhysRevLett.113.077201,
  title = {Quantum Spin Hall Effect in Two-Dimensional Crystals of Transition-Metal Dichalcogenides},
  author = {Cazalilla, M. A. and Ochoa, H. and Guinea, F.},
  journal = {Phys. Rev. Lett.},
  volume = {113},
  issue = {7},
  pages = {077201},
  numpages = {6},
  year = {2014},
  month = {Aug},
  publisher = {American Physical Society},
  doi = {10.1103/PhysRevLett.113.077201},
  url = {https://link.aps.org/doi/10.1103/PhysRevLett.113.077201}
}

@article{Ezawa2012,
  author = {Ezawa, M.},
  title = {Valley-polarized metals and quantum anomalous Hall effect in silicene},
  journal = {Phys. Rev. Lett.},
  volume = {109},
  pages = {055502},
  year = {2012},
    doi = {10.1103/PhysRevLett.109.055502},
  url = { https://doi.org/10.1103/PhysRevLett.109.055502}
}

@article{Xiao2012,
  author = {Xiao, D. and Liu, G.-B. and Feng, W. and Xu, X. and Yao, W.},
  title = {Coupled spin and valley physics in monolayers of MoS2},
  journal = {Phys. Rev. Lett.},
  volume = {108},
  pages = {196802},
  year = {2012},
    doi = {10.1103/PhysRevLett.108.196802},
  url = {https://doi.org/10.1103/PhysRevLett.108.196802}
}

@article{PhysRevLett.110.197402,
  title = {Valley-Spin Polarization in the Magneto-Optical Response of Silicene and Other Similar 2D Crystals},
  author = {Tabert, C. J. and Nicol, E. J.},
  journal = {Phys. Rev. Lett.},
  volume = {110},
  issue = {19},
  pages = {197402},
  numpages = {5},
  year = {2013},
  month = {May},
  publisher = {American Physical Society},
  doi = {10.1103/PhysRevLett.110.197402},
  url = {https://link.aps.org/doi/10.1103/PhysRevLett.110.197402}
}

@article{KaneMele2005a,
  author = {Kane, C. L. and Mele, E. J.},
  title = {Quantum Spin Hall Effect in Graphene},
  journal = {Phys. Rev. Lett.},
  volume = {95},
  pages = {226801},
  year = {2005},
   doi = {10.1103/PhysRevLett.95.226801},
  url = {https://doi.org/10.1103/PhysRevLett.95.226801}
}

@article{KaneMele2005b,
  author = {Kane, C. L. and Mele, E. J.},
  title = {Z2 Topological Order and the Quantum Spin Hall Effect},
  journal = {Phys. Rev. Lett.},
  volume = {95},
  pages = {146802},
  year = {2005}, 
  doi = {10.1103/PhysRevLett.95.146802},
  url = {https://doi.org/10.1103/PhysRevLett.95.146802}
}

@article{nature2014,
  title = { Spin and pseudospins in layered transition metal dichalcogenides },
  author = { Xu, Xiaodong},
  journal = { Nature Physics},
  volume = {10},
  issue = {5},
  pages = {343},
  numpages = {7},
  year = {2014},
  month = {May},
  publisher = {nature},
  doi = {10.1038/nphys2942},
  url = { https://doi.org/10.1038/nphys2942}
}

@article{PhysRevB.101.205408,
  title = {Magneto-optical absorption in silicene and germanene induced by electric and Zeeman fields},
  author = {Muoi, Do and Hieu, Nguyen N. and Nguyen, Chuong V. and Hoi, Bui D. and Nguyen, Hieu V. and Hien, Nguyen D. and Poklonski, Nikolai A. and Kubakaddi, S. S. and Phuc, Huynh V.},
  journal = {Phys. Rev. B},
  volume = {101},
  issue = {20},
  pages = {205408},
  numpages = {12},
  year = {2020},
  month = {May},
  publisher = {American Physical Society},
  doi = {10.1103/PhysRevB.101.205408},
  url = {https://link.aps.org/doi/10.1103/PhysRevB.101.205408}
}

@article{PhysRevB.109.165441,
  title = {Magneto-optical conductivity of monolayer transition metal dichalcogenides in the presence of proximity-induced exchange interaction and external electrical field},
  author = {Li, Y. and Xiao, Y. M. and Xu, W. and Ding, L. and Milo\ifmmode \check{s}\else \v{s}\fi{}evi\ifmmode \acute{c}\else \'{c}\fi{}, M. V. and Peeters, F. M.},
  journal = {Phys. Rev. B},
  volume = {109},
  issue = {16},
  pages = {165441},
  numpages = {14},
  year = {2024},
  month = {Apr},
  publisher = {American Physical Society},
  doi = {10.1103/PhysRevB.109.165441},
  url = {https://link.aps.org/doi/10.1103/PhysRevB.109.165441}
}

@article{PhysRevB.101.045424,
  title = {Magneto-optical transport properties of monolayer transition metal dichalcogenides},
  author = {Hien, Nguyen D. and Nguyen, Chuong V. and Hieu, Nguyen N. and Kubakaddi, S. S. and Duque, C. A. and Mora-Ramos, M. E. and Dinh, Le and Bich, Tran N. and Phuc, Huynh V.},
  journal = {Phys. Rev. B},
  volume = {101},
  issue = {4},
  pages = {045424},
  numpages = {13},
  year = {2020},
  month = {Jan},
  publisher = {American Physical Society},
  doi = {10.1103/PhysRevB.101.045424},
  url = {https://link.aps.org/doi/10.1103/PhysRevB.101.045424}
}

@article{Ezawa_2012,
doi = {10.1088/1367-2630/14/3/033003},
url = {https://doi.org/10.1088/1367-2630/14/3/033003},
year = {2012},
month = {mar},
publisher = {IOP Publishing},
volume = {14},
number = {3},
pages = {033003},
author = {Ezawa, Motohiko},
title = {A topological insulator and helical zero mode in silicene under an inhomogeneous electric field},
journal = {New Journal of Physics}
}

@article{Ezawa2012a,
doi = {10.1209/0295-5075/98/67001},
url = {https://doi.org/10.1209/0295-5075/98/67001},
year = {2012},
month = {jun},
publisher = {},
volume = {98},
number = {6},
pages = {67001},
author = {Ezawa, Motohiko},
title = {Dirac theory and topological phases of silicon nanotube},
journal = {Europhysics Letters},
}

@article{PhysRevB.93.035406,
  title = {Quantum magnetotransport properties of a ${\text{MoS}}_{2}$ monolayer},
  author = {Tahir, M. and Vasilopoulos, P. and Peeters, F. M.},
  journal = {Phys. Rev. B},
  volume = {93},
  issue = {3},
  pages = {035406},
  numpages = {9},
  year = {2016},
  month = {Jan},
  publisher = {American Physical Society},
  doi = {10.1103/PhysRevB.93.035406},
  url = {https://link.aps.org/doi/10.1103/PhysRevB.93.035406}
}

@article{PhysRevB.94.045415,
  title = {Magneto-optical transport properties of monolayer ${\mathrm{WSe}}_{2}$},
  author = {Tahir, M. and Vasilopoulos, P.},
  journal = {Phys. Rev. B},
  volume = {94},
  issue = {4},
  pages = {045415},
  numpages = {8},
  year = {2016},
  month = {Jul},
  publisher = {American Physical Society},
  doi = {10.1103/PhysRevB.94.045415},
  url = {https://link.aps.org/doi/10.1103/PhysRevB.94.045415}
}

@article{PhysRevB.92.125303,
  title = {Linear magnetotransport in monolayer ${\mathrm{MoS}}_{2}$},
  author = {Wang, C. M. and Lei, X. L.},
  journal = {Phys. Rev. B},
  volume = {92},
  issue = {12},
  pages = {125303},
  numpages = {10},
  year = {2015},
  month = {Sep},
  publisher = {American Physical Society},
  doi = {10.1103/PhysRevB.92.125303},
  url = {https://link.aps.org/doi/10.1103/PhysRevB.92.125303}
}
\end{document}